%% file: main.tex
\documentclass[a4paper,11pt]{article}
\usepackage{jheppub}

\pdfoutput=1
\usepackage{amsmath}
\usepackage{gensymb}
\usepackage{amsfonts}
\usepackage{comment}
\usepackage{amssymb}
\usepackage{mathrsfs}
\usepackage{graphicx}
\usepackage{color}
\usepackage{multirow}
\usepackage{bm}
\usepackage{tabularx}
\usepackage{subcaption}
\usepackage{blindtext}
\usepackage{wasysym}
\usepackage{makecell} % For line breaks in table cells
\usepackage{float}
\usepackage{orcidlink}
\usepackage[normalem]{ulem}
\usepackage{lineno}
\usepackage[T1]{fontenc}
\usepackage{soul}
\usepackage{colortbl}
\usepackage{array}
\usepackage{cancel}
\usepackage[utf8]{inputenc}
\newcolumntype{L}{>{\centering\arraybackslash}m{1.99cm}}
\newcommand{\bi}{\begin{itemize}}
\newcommand{\ei}{\end{itemize}}

\makeatletter
\gdef\@fpheader{}
\makeatother

\begin{document}
 
\title{Feasibility study of light sterile neutrino searches with a future NINJA-like detector}
%\\ 
%\vspace{4mm}
%{\small (The NINJA Collaboration)}}

\input{authors.tex}
\authorlist

\abstract{
In this paper, we investigate the sensitivity of a future NINJA-like experiment at J-PARC to eV-scale sterile neutrinos within the 3+1 framework. We perform a phenomenological feasibility study using the $\nu_\mu \rightarrow \nu_e$ appearance, $\nu_\mu \rightarrow \nu_\mu$ and $\nu_e \rightarrow \nu_e$ disappearance channels, focusing on possible future configurations of a detector located at different floors of the NM building (B2, SS, and GROUND), corresponding to different off-axis angles. Our analysis is based on a simplified and effective detector response, in which events are classified into electron-like and muon-like topologies and constant benchmark selection efficiencies are applied. We explore different exposure scenarios and assess the impact of analysis choices such as upper energy cuts. We include systematic uncertainties on signal and background rates, and study the robustness of our results with respect to variations in the assumed energy resolution, and vary efficiencies for key backgrounds such as muon misidentification from charge-current and neutral-current interactions. Finally, we examine the effects of combining data from multiple detector locations. We find that the SS floor provides the strongest constraints on the active–sterile mixing parameters, while the B2 and GROUND configurations offer constraints comparable to current bounds for the probed mass-squared differences. Our results indicate that a NINJA-like detector, optimized for sufficient statistics and benchmark identification performance, has the potential to provide competitive constraints on light sterile neutrino scenarios in future runs.}

\maketitle

\section{Introduction}
Since the initial observations of anomalies in neutrino oscillation experiments, such as the significant excesses of electron antineutrino events detected at low energies in both the pion decay-at-rest (LSND~\cite{LSND:2001aii}) and pion decay-in-flight (MiniBooNE~\cite{MiniBooNE:2020pnu}) experiments, extensive theoretical and experimental efforts have been undertaken over the past few decades to interpret and confirm these unexpected results~\cite{Acero:2022wqg}. The most straightforward explanation for these deviations from the standard three-neutrino oscillation behavior involves eV-scale sterile neutrinos, which can mix with active neutrinos and induce short-baseline oscillations. Sterile neutrinos appear naturally in many extensions of the Standard Model (for example see Refs.~\cite{Dvali:1998qy,Bringmann:2022aim,Chun:1995js,Hebbar:2017fit,Barry:2011wb,Zhang:2011vh}). The simplest such model extends the framework to a so-called 3+1 model by introducing a single sterile neutrino. Building on the original LSND and MiniBooNE anomalies, a combined analysis of MiniBooNE and MicroBooNE data further explored the 3+1 sterile neutrino hypothesis and found it to remain statistically viable~\cite{MiniBooNE:2022emn}. More recently, MicroBooNE has presented updated results that place strong constraints on sterile-neutrino interpretations of the MiniBooNE low-energy excess \cite{MicroBooNE:2025nll}. These findings significantly restrict specific regions of the sterile neutrino parameter space associated with that anomaly, but do not exclude all short-baseline sterile neutrino scenarios motivated by accelerator-based experiments. This highlights the continued importance of complementary probes by using different baselines, energies, and detector technologies. Beyond accelerator experiments, the three-neutrino framework has also been challenged by deficits in measured $\nu_e$ rates from calibrated sources in GALLEX~\cite{Kaether_2010} and BEST~\cite{Barinov:2021asz}, as well as by results from the Neutrino-4 reactor antineutrino experiment~\cite{Serebrov:2020kmd}, which suggested oscillations over a baseline of several meters. However, data from experiments such as IceCube \cite{IceCube:2024nle}, Daya Bay-MINOS \cite{MINOS:2020iqj}, T2K \cite{T2K:2019efw}, NO$\nu$A \cite{NOvA:2024imi}, STEREO-PROSPECT-DANSS \cite{Stereo:2021wfd}, NEOS \cite{NEOS:2016wee}, DeepCore \cite{Abbasi:2024ktc}, ICARUS \cite{ICARUS:2026lvh} and JSNS$^2$ \cite{JSNS2:2026muy} are consistent with the no-sterile-neutrino hypothesis. Several upcoming experiments, including ESSnuSB \cite{Ghosh:2019zvl}, DUNE \cite{DUNE:2020fgq}, nuSCOPE \cite{Blanchet:2026jwx} and SBND \cite{Machado:2019oxb} at Fermilab's short-baseline program are dedicated to the search for sterile neutrinos. Given the conflicting indications from different experiments and the statistical significance of their results, the existence of a light sterile neutrino remains an open question, which needs further investigation from multiple perspectives. With this in mind, in this paper, we study the capability of the NINJA-like detector to test the sterile neutrino hypothesis and constrain its parameters during its future runs. Although NINJA’s primary physics objective is to measure neutrino–nucleus cross sections {\color{red} \cite{NINJA:2020bvx}}, its proximity to the neutrino source at J-PARC (280~m baseline), 
high-resolution tracking capability, and effective suppression of neutral current (NC) $\pi^0$ backgrounds via emulsion vertex reconstruction make it a well-suited and 
cost-effective probe of eV-scale sterile neutrino oscillations at short baselines. Here we would like to point out that though the “vanilla” 3+1 model often clashes with the  cosmological measurements \cite{Elbers:2025vlz} and suffers from tension between the appearance and disappearance channel measurements, it is still a community standard for comparing performance across different experiments. Although in this work, we mainly focus on the 3+1 scenario, the NINJA setup can also be used to study models with heavy neutral leptons \cite{MicroBooNE:2022ctm}, dark sectors \cite{MicroBooNE:2025khi}, decaying sterile neutrinos \cite{Dentler:2019dhz} etc.

The paper is organized as follows: We outline the theoretical framework for the sterile neutrino hypothesis in section \ref{sec:formalism}, describe the experimental configuration of NINJA-like considered in our analysis in section \ref{sec:simulation}, and present our findings in section \ref{sec:res}. Finally, in section \ref{sec:conclusion} we summarize our results and conclude.

\section{Theoretical formalism}
\label{sec:formalism}
In the simplest extension of the Standard Model with one additional neutrino state that is not gauged under weak interactions, oscillation physics remains very similar to the case with three active neutrinos. In this scenario, an extra mass eigenstate, $m_4$ (assumed to be much heavier than the active ones), and a corresponding flavor eigenstate, $\nu_s$, are added to the standard framework. Without introducing parametrization dependence, the leptonic mixing matrix is typically extended to a $4\times4$ form as follows:

\begin{equation}
    V = U_{34}(\theta_{34}, \delta_{34}) U_{24}(\theta_{24}, \delta_{24}) U_{14}(\theta_{14}, 0) V_{3\nu},
\end{equation}
where
\begin{equation}
    V_{3\nu} = U_{23}(\theta_{23}, 0) U_{13}(\theta_{13}, \delta_{CP}) U_{12}(\theta_{12}, 0),
\end{equation}
and $U_{ij}(\theta_{ij}, \delta_{ij})$ denotes a rotation in the $(i, j)$ plane with mixing angle $\theta_{ij}$ and phase $\delta_{ij}$.

Adding an extra sterile neutrino increases the number of parameters by introducing an additional mass-squared difference, $\Delta m^2_{41}$, three new mixing angles ($\theta_{14}$, $\theta_{24}$, and $\theta_{34}$), and two additional CP-violating phases ($\delta_{24}$ and $\delta_{34}$). However, these phases only lead to an observable CP violation if there is interference between oscillations driven by different mass-squared differences - if both $\Delta m^2_{41}$ and one of the standard splittings ($\Delta m^2_{21}$ or $\Delta m^2_{31}$) contribute to the oscillation. 
In the short-baseline (SBL) approximation, we assume 
\begin{equation}
\Delta m^2_{41} \approx \Delta m^2_{42} \approx \Delta m^2_{43} \quad \text{and} \quad \Delta m^2_{21} \approx \Delta m^2_{32} \approx 0,
\end{equation}
which means that oscillations due to solar and atmospheric mass splittings are suppressed. Consequently, a new large mass-splitting allows for SBL oscillations. Under these conditions, the probabilities that are relevant for experimental searches are approximated by \cite{Acero:2022wqg}
{\small
\begin{align}
P(\nu_e \to \nu_e) &\simeq 1 - 4\,|V_{e4}|^2\,(1 - |V_{e4}|^2)\,\sin^2\!\left(\frac{\Delta m^2_{41}\,L}{4\,E_\nu}\right) = 1 - \sin^2\!\left(2\theta_{ee}\right)\,\sin^2\!\left(\frac{\Delta m^2_{41}\,L}{4\,E_\nu}\right), \label{eq:pee} \\[3.5pt]
P(\nu_\mu \to \nu_\mu) &\simeq 1 - 4\,|V_{\mu4}|^2\,(1 - |V_{\mu4}|^2)\,\sin^2\!\left(\frac{\Delta m^2_{41}\,L}{4\,E_\nu}\right) = 1 - \sin^2\!\left(2\theta_{\mu\mu}\right)\,\sin^2\!\left(\frac{\Delta m^2_{41}\,L}{4\,E_\nu}\right), \label{eq:pmu} \\[3.5pt]
P(\nu_\mu \to \nu_e) &\simeq 4\,|V_{\mu4}|^2\,|V_{e4}|^2\,\sin^2\!\left(\frac{\Delta m^2_{41}\,L}{4\,E_\nu}\right)=\sin^2\!\left(2\theta_{\mu e}\right)\,\sin^2\!\left(\frac{\Delta m^2_{41}\,L}{4\,E_\nu}\right). \label{eq:pemu}
\end{align}
}
where we defined effective mixing angles that capture the amplitude of oscillations and are parametrization independent: 
\begin{align}\label{eq:oscillation}
    \sin^2 2\theta_{ee} \equiv 4\,|V_{e4}|^2\,(1 - |V_{e4}|^2), \\
\sin^2 2\theta_{\mu\mu} \equiv 4\,|V_{\mu4}|^2\,(1 - |V_{\mu4}|^2), \\
\sin^2 2\theta_{\mu e} \equiv 4\,|V_{\mu4}|^2\,|V_{e4}|^2. \label{eq:oscillation_1}
\end{align}
A key observation is that if both \(|V_{e4}|^2\) and \(|V_{\mu4}|^2\) are nonzero, then one expects to observe \(\nu_e\) disappearance, \(\nu_\mu\) disappearance, and \(\nu_e\) appearance. By examining equations \eqref{eq:pee}-\eqref{eq:pemu}, we see that for a 3+1 model with \(\Delta m_{41}^2 \approx 1\,\mathrm{eV}^2\), motivated by LSND data \cite{LSND:2001aii}, the oscillation effects become significant for \(L/E_\nu\) values such that
\begin{align}
\frac{\Delta m_{41}^2\,L}{4E_\nu} \sim \mathcal{O}(1).
\end{align}
This implies that \(L\) is in km (or m) when \(E_\nu\) is in GeV (or MeV), which sets the scale for SBL experiments. Note that at this baseline, the 3+1 scenario is basically a two flavour case. So naturally, the CP violation phases does not appear in the probability, which can be seen via Eqs. \ref{eq:oscillation} - \ref{eq:oscillation_1}.

\section{Assumptions used for the future NINJA-like detector}
\label{sec:simulation}

The NINJA (\textbf{N}eutrino \textbf{I}nteraction research with \textbf{N}uclear emulsion and \textbf{J}-PARC \textbf{A}ccelerator) experiment is designed to precisely measure the cross sections of neutrino-nucleus interactions using emulsion cloud chamber (ECC) technology \cite{NINJA:2020gbg}. The main detector technology used in NINJA is an Emulsion Cloud Chamber (ECC), which consists of alternating layers of tracking films and interaction targets. In a current detector configuration, each tracking layer consists of two 350 $\mu m$-thick emulsion films coated with an emulsion gel, placed on both sides of a 500 $\mu m$ iron plate, which is used for momentum measurements. This layered design allows one to reconstruct detailed neutrino interaction events with high spatial resolution. 

\begin{figure}[H]
\centering
\begin{subfigure}[t]{0.45\textwidth}
\centering
\includegraphics[width=\textwidth]{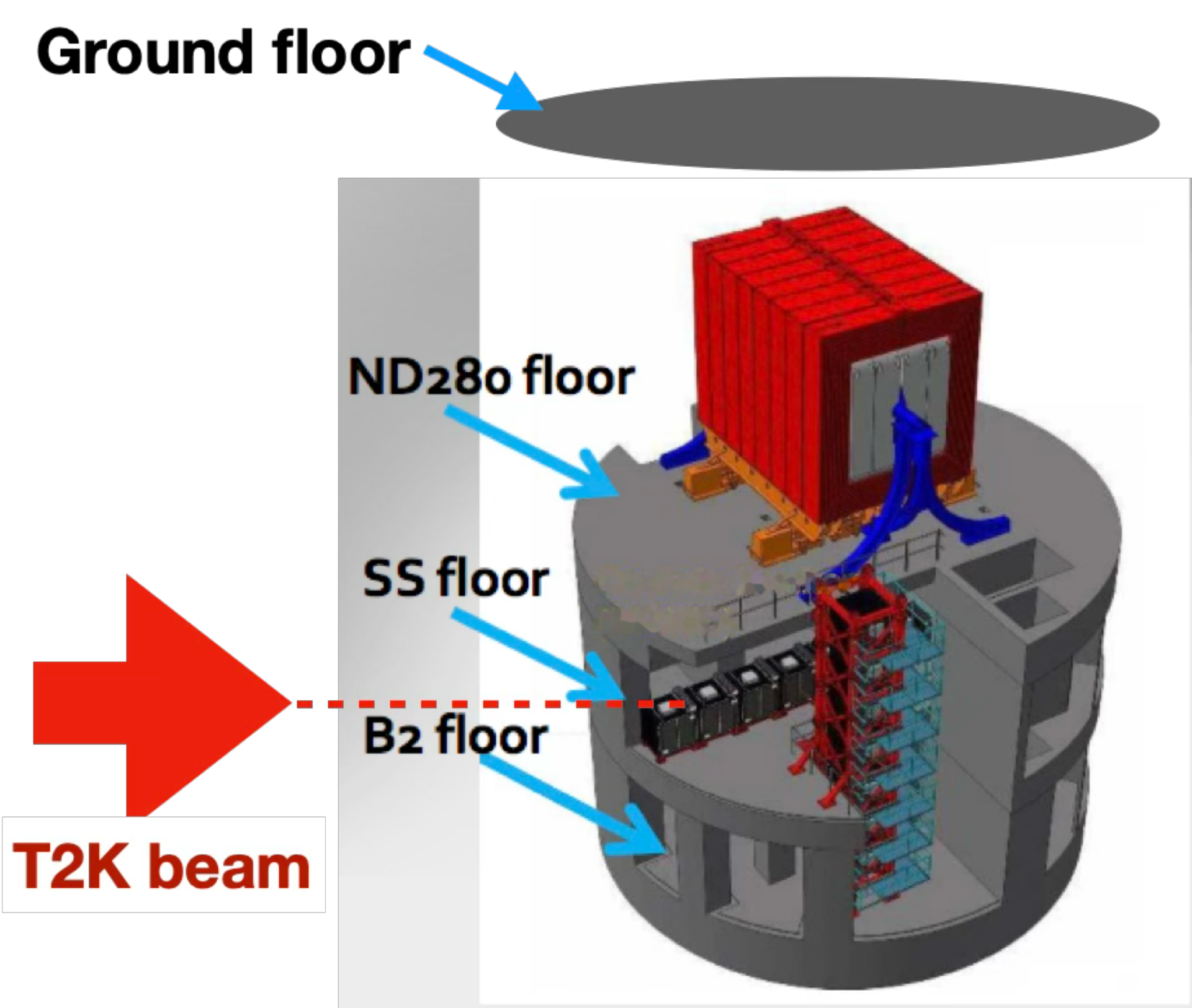}
\caption{Location of the detector within the NM building at J-PARC, showing the B2 (1.5$^\circ$ off-axis), SS (on-axis), and GROUND (6$^\circ$ off-axis) floors considered in this analysis. Adapted from the T2K technical design report~\cite{T2K:2019bbb}.}
\label{fig:nm_building}
\end{subfigure}
\hfill
\begin{subfigure}[t]{0.45\textwidth}
\centering
\raisebox{0.5cm}{\includegraphics[width=\textwidth]{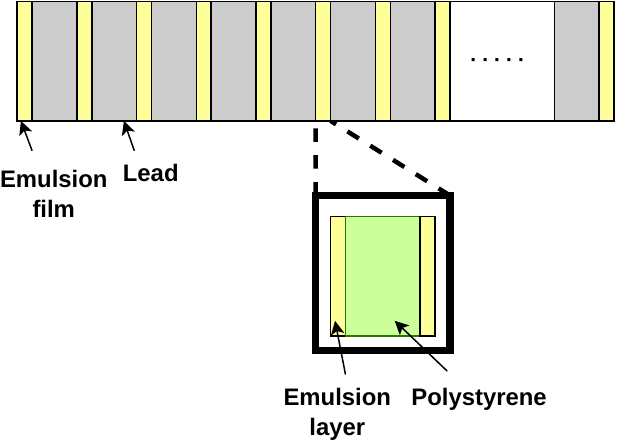}}
\caption{Schematic of a single ECC tracking unit. The ECC consists of emulsion films interleaved with lead plates, used for muon momentum reconstruction via multiple Coulomb scattering; each film comprises a polystyrene base coated with emulsion layers on both sides. Lead, polystyrene, and emulsion all act as neutrino targets.}
\label{fig:ecc_layer}
\end{subfigure}
\caption{Illustration of the envisioned NINJA-like detector geometry
and its placement relative to the beamline.}
\label{fig:detector_geometry}
\end{figure}

The aim of this work is to perform a phenomenological feasibility and sensitivity analysis for probing eV-scale sterile neutrinos with a future NINJA-like detector configuration. In particular, we explore the experimental conditions and detector performance requirements under which such a measurement could be competitive. In the current runs of NINJA, the detector is located at the \textbf{B2} floor (1.5$^\circ$ off-axis) of the NM building within the J-PARC facility, shown in Fig.~\ref{fig:nm_building}. There is also a possibility to place the NINJA detector at the \textbf{SS} floor (on-axis), and \textbf{GROUND} floor (6$^\circ$ off-axis) of the same building in its future runs. In our analysis, we considered these three detector locations, assuming a 280~m distance from the neutrino production point at J-PARC. The flux corresponds to $10^{21}$ POT (Protons On Target) per year. While the current NINJA detector employs water and iron as the target material, future configurations may consider denser materials such as lead in order to achieve sufficient fiducial mass for short-baseline oscillation studies. In the present analysis, we assume a lead-based ECC target, with the layered structure illustrated in Fig.~\ref{fig:ecc_layer}, and explore different exposures in units of ton-year in the neutrino mode.

 A detailed detector simulation for such a future lead-based NINJA-like configuration is not yet available. Therefore, reconstruction performance, particle identification, and background rejection are not explicitly modeled. Instead, detector effects are incorporated through a simplified and effective description. Events are classified into electron-like and muon-like candidate topologies, and constant selection efficiencies are applied to these categories. These efficiencies should be interpreted as benchmark performance requirements, motivated by the known capabilities of ECC technology demonstrated in previous neutrino experiments and benchmark values from other accelerator based experiments \cite{Arrabito:2007rq,Alekou:2022emd,NINJA:2022zbi} rather than as predictions for NINJA performance. Backgrounds are treated in two categories. The first category consists of irreducible backgrounds - the intrinsic $\nu_e$ and $\bar{\nu}_e$ components of the $\nu_\mu$ beam and wrong-sign contributions, which are experimentally indistinguishable from oscillation-induced signal events at the topology level and therefore enter the reconstructed samples with 100\% efficiency. The $\nu_e$ component originates primarily from kaon and muon decays in the decay pipe. The second category consists of misidentified (mis-ID) charge current (CC) and neutral current (NC) events. For $\nu_\mu$ CC events, muon identification in the NINJA experiment relies on the downstream Baby MIND (Magnetized Iron Neutrino Detector) spectrometer, which tags muons leaving the ECC target and provides their charge and momentum~\cite{Odagawa:2022crm}. However, low-momentum muons that stop within the ECC before reaching Baby MIND may be reconstructed as electron-like events, since their short tracks can mimic the early stages of an electromagnetic (EM) shower. The probability of this occurring depends on the muon momentum and the ECC thickness. In the absence of a dedicated simulation for this effect, we adopt 1\% as a benchmark mis-ID efficiency for $\nu_\mu$ CC events, with 0.5\% as an optimistic alternative, to be validated by future NINJA runs. For NC events, the dominant misidentification mechanism is decay $\pi^0 \to \gamma\gamma$, where the two photons convert in the lead and produce overlapping electromagnetic showers that mimic a single electron. In an ECC detector, the thickness of the plates plays a critical role in the electron identification and background rejection. In particular, thinner lead plates reduce the probability of early photon conversion and improve the separation between single-electron EM showers and backgrounds originating from NC interactions. Conversely, excessively thick plates can increase the likelihood of photon conversions in the first layer, thereby enhancing NC-induced backgrounds. In the absence of a dedicated detector simulation for a future lead-based NINJA-like configuration, we do not fix the thickness of the lead plate or the total radiation length of the ECC stack in this study. Instead, we assume that the detector geometry would be optimized to achieve the benchmark efficiencies for electron-like and muon-like identification adopted in the analysis. In this work, we adopt a NC efficiency benchmark of 0.5\%, with an alternative value of 1\%. Signal events are assumed to be reconstructed with 95\% efficiency.

Our assumption of a mis-ID background of 0.5\% - 1\% can be validated in the following way. The NC mis-ID efficiency can be estimated as: 
\begin{align}
{\rm multiplicity~of}~\pi^0 \times 2 \gamma \times {\rm probability~of}~\gamma \rightarrow e^{\pm}~{\rm pair~production} \\ \nonumber 
~\times~{\rm probability~of~misidentification~of~electron~and~positron~as~one~electron}. 
\end{align}
For the 1 GeV energy region in water, the ratio of the relative cross section for NC1$\pi^0$ interactions to the total $\nu_\mu$ CC cross section is measured to be 0.064 $\pm$ 0.001 (stat.) $\pm$ 0.007 (sys.) \cite{K2K:2004qpv}. Taking into effect the final state interactions, this results in a multiplicity $\pi^0$ being $0.06 - 0.08$. In lead, this number falls below $0.05$, as 80\% of pions are absorbed inside the nucleus. The probability of pair production $\gamma \rightarrow e^{\pm}$ after traversing thickness $x$ can be computed using the standard exponential attenuation law 
\begin{equation}
P(x)=1-\exp\left(-\frac{x}{\lambda_{\mathrm{pair}}}\right) 
\end{equation}
with constant pair production mean free path $\lambda_{\mathrm{pair}}=9X_{0}/7$, where $X_{0}$ is the radiation length of the lead that can be obtained from the particle data group (PDG) 
\cite{ParticleDataGroup:2024cfk}. Assuming 1~mm thickness, this probability is around 10\%. Finally, assuming a track detection efficiency of $97\%$ in nuclear emulsion \cite{NINJA:2020gbg}, the probability of misidentifying an electron-positron pair as one electron is $0.97 \times 0.03$. Putting all the numbers together, we see that the efficiency of NC mis-ID is significantly below 1\%. Note that unlike NC efficiencies, CC efficiencies can not be estimated empirically and therefore should be estimated by detailed simulation. As electron neutrino efficiencies for an emulsion detector in the low energy region has not been computed yet, as mentioned earlier, our values (95\% for signal and 0.5\% - 1\% for background) can be considered as benchmark performance requirements and inspired from ECC technology demonstrated in previous neutrino experiments and other accelerator based experiments \cite{Arrabito:2007rq,Alekou:2022emd,NINJA:2022zbi}.

In the analysis, we implement the detector performance using GLoBES~\cite{Huber:2004ka, Huber:2007ji}, which requires the efficiencies to  be specified for the individual channels. We therefore assign topology-based efficiencies to each channel as an effective implementation, encoding how different physical processes populate the reconstructed electron-like or muon-like samples. The $\chi^2$ is evaluated using a Poissonian log-likelihood over spectral bins with a normalization pull term. 
The data set is generated according to the standard three-flavor hypothesis without sterile mixing and the scan is performed on $\Delta m^2_{41}$ and $\sin^2 2\theta_{\mu e}$ when both electron appearance and muon disappearance channels are considered and on $\Delta m^2_{41}$ and $\sin^2 2\theta_{\mu \mu}$ when only muon disappearance channel is considered. The energy range is 0-10 GeV, divided into 100 uniform bins. The exclusion curves are drawn at $\Delta \chi = 5.99$, corresponding to 95\% C.L. in the 2-dimensional parameter plane. We estimate the sensitivity by combining the \(\nu_e\) disappearance, \(\nu_\mu\) disappearance, and \(\nu_e\) appearance channels. However, the main sensitivity comes from the \(\nu_e\) appearance and the \(\nu_\mu\) disappearance channel. 

Neutrino energy reconstruction is modeled using a Gaussian energy resolution with a relative width of 10\%, applied uniformly to all channels. Note that in the absence of a realistic smearing matrix, approximating the event reconstruction method with a Gaussian energy resolution performs quite well as shown in Ref.~\cite{Chatterjee:2021wac} in the context of DUNE. This provides an effective description of the reconstruction effects in the absence of detailed detector modeling. However, one of the drawbacks can be the case when energy being carried away in the form of neutrons, as this could lead to an asymmetric tail in the reconstructed energy distribution. In our case, since we have narrow flux for the GROUND floor and the B2 floor, this effect is expected to be small. For SS floor, this effect can be important and there might be some change of sensitivity if one considers a realistic smearing. Our benchmark value of 10\% is motivated both by recent NINJA results showing that muons with momenta up to approximately 1 GeV/c achieve a momentum resolution of 10–30\% \cite{NINJA:2026pme} and previous studies in the context of long-baseline experiments \cite{Alekou:2022emd}. However, later we will show that the impact of the energy resolution on our results is negligible. 

Systematic uncertainties are included through overall normalization uncertainties of 5\% for signal and 10\% for background, unless otherwise mentioned. The justification for our choice is as follows. The systematic uncertainties in the NINJA-like experiment come from mainly uncertainty in flux measurement, uncertainty in cross-section measurement, and detector related uncertainties. Recent analysis of NINJA data shows that uncertainties associated with beam flux, neutrino interaction, detector response, are relatively stable at approximately 5\%, 10–15\% and 10\%, respectively \cite{NINJA:2026pme}. Further, as our setup will use beam from the J-PARC facility, and the energy region is similar to that of Hyper-K, we can safely consider the systematics predicted by the T2HK experiment as our benchmark. According to \cite{Hyper-Kamiokande:2025fci}, the current T2K systematics corresponds to an uncertainty of 4.7\% combining uncertainties coming from flux, cross-section and detector for one ring electron-like events. For T2HK, an improved error model considering 10 years of data accumulated at the near and far detectors will bring this systematic to 2.1\% and NC systematics to 10\%. For reference, the DUNE technical design report assumes a 2\% systematic error on the signal, and a 5\% error on the background \cite{DUNE:2021cuw}. Given this, our choice of 5\% signal error and 10\% background error for a future NINJA-like detector is realistic. Nevertheless, later we will show that our results are still statistically limited and therefore have minimal dependence on systematics. Irrespective of the above discussion, we would like to state that our sensitivities are idealistic in the sense that they do not include a detailed treatment of systematic uncertainties which necessitates follow-up studies with full MC simulation. Therefore, comparison of our sensitivity with other experiments, for example MicrooBooNE, should be done with caution as the sophistication of the systematics treatment is very different between them.

\section{Results}
\label{sec:res}
{\it Probability, sensitivity, and event rates} \\

\begin{figure}[h!]
\begin{center}  
\includegraphics[width=118mm, height=75mm]{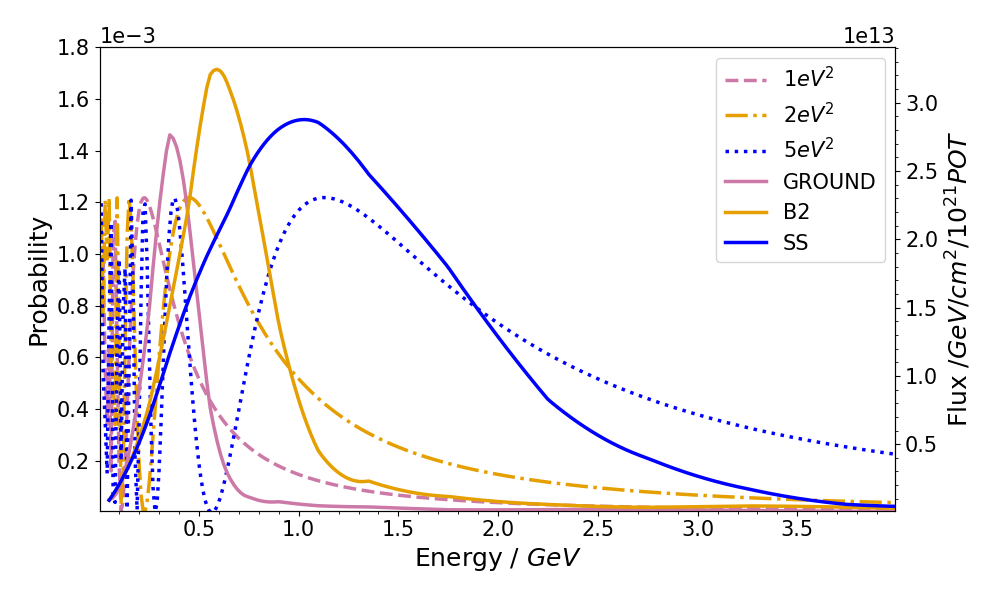}
\caption{Probability (left), presented with dashed lines, and $\nu_\mu$ flux (right), presented with solid lines, relevant for NINJA.}
\label{fig_prob}
\end{center}
\end{figure}

\begin{figure}[h!]
\centering
\begin{subfigure}[t]{0.49\textwidth}
    \centering
    \includegraphics[width=\textwidth,height=65mm]{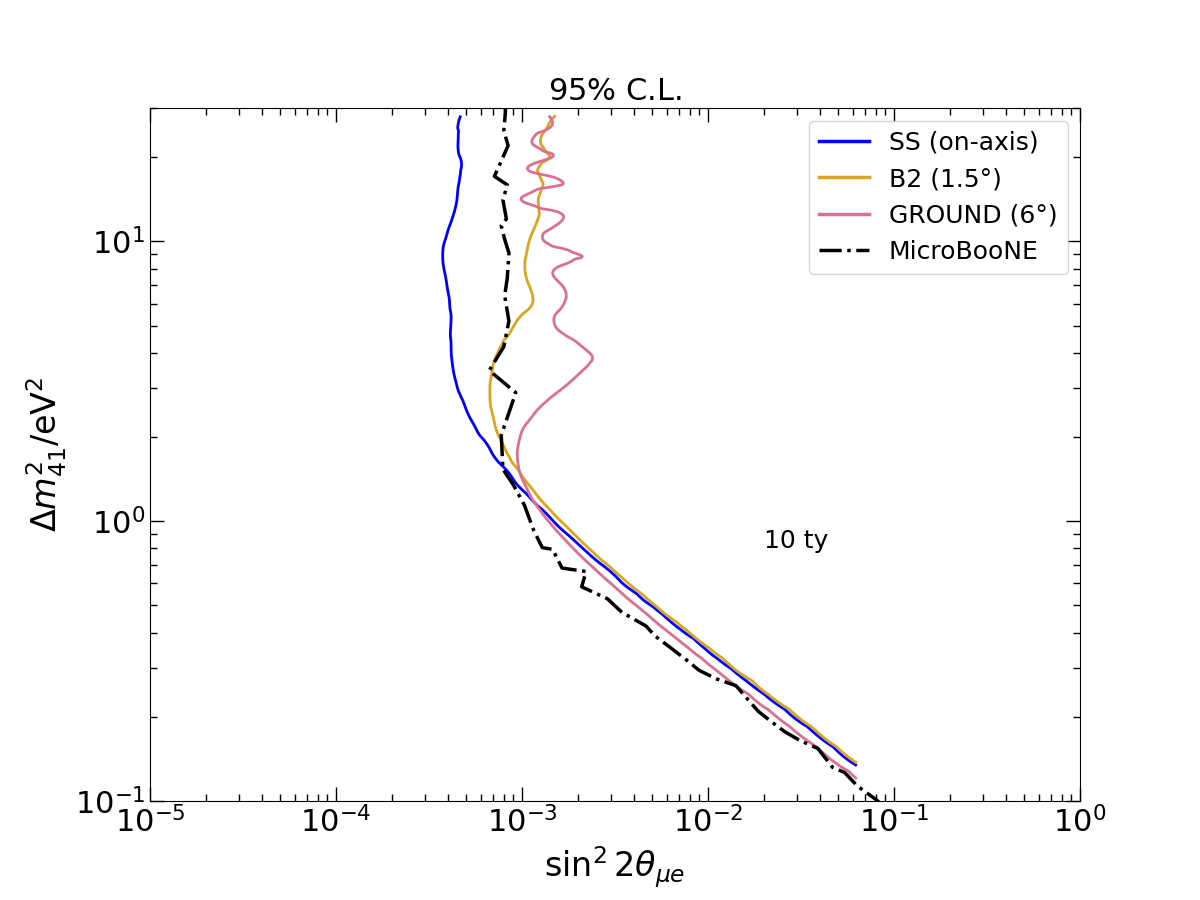}
    \caption{Sensitivity for three detector locations with 10 ton-year exposure.}
    \label{fig:sens_locations}
\end{subfigure}%
\hfill
\begin{subfigure}[t]{0.49\textwidth}
    \centering
    \includegraphics[width=\textwidth, height=65mm]{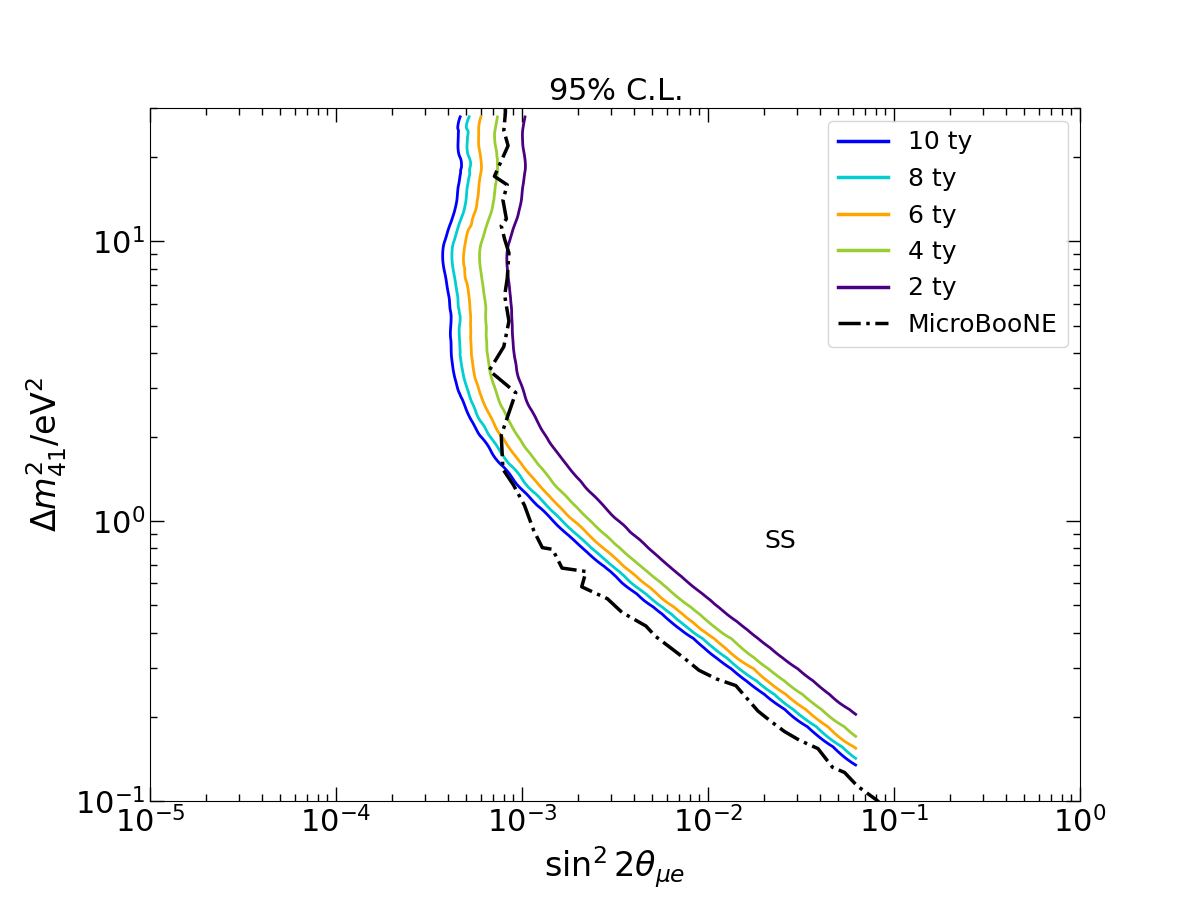}
    \caption{Sensitivity for various exposures at the SS floor.}
    \label{fig:sens_exposures}
\end{subfigure}
\caption{Bounds on the sterile mixing parameters.}
\label{fig_sens}
\end{figure}

In Fig.~\ref{fig_prob}, we show the appearance channel probability for sterile neutrinos as a function of the neutrino energy for various mass-square differences (dashed curves). The muon-neutrino fluxes at the three detector positions are depicted by solid curves, with $\sin^22\theta_{\mu e}$ = 0.001, since NINJA is sensitive in this region. We observe that if the detector is located on floor B2, NINJA will be sensitive to a mass-square difference of 2 eV$^2$, as probability and flux peaks coincide in the same energy region. Similarly, the SS floor is sensitive to 5 eV$^2$, and the GROUND floor to 1 eV$^2$. 
In the left panel of Fig.~\ref{fig_sens} we show the expected sensitivity of future NINJA-like detector to constrain the sterile mixing parameters in the $\sin^2 2\theta_{\mu e}$–$\Delta m^2_{41}$ plane at 95\% C.L \footnote{In our calculation we assumed that all the neutrinos are produced at the target. We have discussed the effect of decay pipe in Appendix A.}. Three detector positions are compared using a 10 ton-year exposure. For comparison, we have also included the current limit from the MicroBooNE experiment \cite{MicroBooNE:2025nll} in this panel\footnote{In the Appendix B, we have shown the comparison of our results with the combined expected sensitivity of the short-baseline neutrino (SBN) program at Fermilab.}. As anticipated by the flux–probability overlap in Fig.~\ref{fig_prob}, this panel confirms the sensitivity of the different flux options to different values of $\Delta m^2_{41}$. From the left panel of Fig.~\ref{fig_sens} we see that with the SS floor configuration, future NINJA-like detector can place stronger constraints on sterile mixing parameters than the current MicroBooNE limit, while the GROUND floor yields a weaker bound and the B2 floor is comparable to the existing bound. The pink curve, which corresponds to the GROUND detector location, exhibits oscillatory effects at higher $\Delta m^2_{41}$.
This is because its low‑energy, narrowly peaked flux ($E\sim0.3\text{–}0.4$~GeV) is sensitive to multiple values of $\Delta m^2_{41}$ and spans several oscillation cycles of
\begin{equation}
\phi(E) = \frac{\Delta m^2_{41}\,L}{4E},\,
\end{equation}
\noindent
which is the argument of the $\sin^2$ term entering the oscillation probabilities in Eqs.~\eqref{eq:pee}--\eqref{eq:pemu}. Since $\sin^2(\phi)$ has period $\pi$, each increase of $\phi$ by $\pi$ corresponds to one oscillation cycle - the interval between two probability maxima (or minima) as a function of energy. For high values of $\Delta m^2_{41}$, these $\pi$-intervals map onto closely spaced energy intervals, so the oscillation maxima are closely spaced in energy. Because the GROUND flux is narrowly peaked at low energy ($E \sim 0.3$--$0.4$~GeV), it overlaps with several closely spaced maxima for the $\Delta m^2_{41}$ values probed here, producing the rapid oscillatory structure visible in Fig.~\ref{fig_prob}, which is not the case for SS and B2 floor flux.  These oscillations remain visible with our energy resolution. Since the SS floor offers the best sensitivity to sterile neutrinos, the right panel of Fig.~\ref{fig_sens} shows the sensitivity achieved on the SS floor for varying exposures. We find that, even for a relatively modest exposure of 4-6 ton-year, the projected sensitivity becomes comparable to, and in some regions slightly stronger than the existing constraints of the MicroBooNE, particularly around $\Delta m^2_{41} \sim 5~\mathrm{eV}^2$. 

The trend in overall sensitivity becomes clear upon examining 
tables~\ref{tab:appearance} and~\ref{tab:disappearance}, which list the numbers 
of signal and major background events at each detector location for the $\nu_e$ appearance channel and $\nu_\mu$ disappearance channel, respectively. The count of 
signal events on the SS floor exceeds those on the B2 and GROUND floor, 
reflecting the higher on-axis flux. The efficiency conventions and background 
definitions follow Sec.~\ref{sec:simulation}; numbers outside (inside) 
parentheses correspond to the benchmark (alternative) efficiency values. For the $\nu_e$ appearance channel (table~\ref{tab:appearance}), the dominant background is the intrinsic $\nu_e$ beam component, followed by misidentified $\nu_\mu$ CC and NC events, and the intrinsic $\bar{\nu}_e$ wrong sign contribution. For the $\nu_\mu$ disappearance channel (table~\ref{tab:disappearance}), the dominant background is the intrinsic wrong sign $\bar{\nu}_\mu$ beam component, originating from negatively charged pions and kaons not fully suppressed by the horn focusing system, which is experimentally indistinguishable from the signal at the topology level. For the $\nu_e$ disappearance channel, the intrinsic $\nu_e$ beam in table~\ref{tab:appearance} now becomes the signal, with the dominant backgrounds being misidentified $\nu_\mu$ CC and NC events reconstructed as electron-like. 

\begin{table}
\centering
\caption{Expected background and signal event counts for the $\nu_e$ \textbf{appearance} channel, for 10 ton-year exposure at each detector location. Numbers outside (inside) parentheses correspond to the benchmark (alternative) mis-ID and NC efficiencies of 1\% and 0.5\% (0.5\% and 1\%), respectively. Events are generated with $\sin^2 2\theta_{\mu e} = 0.001$.}
\label{tab:appearance}
\begin{tabular}{llccc}
\hline\hline
\multicolumn{2}{l}{} 
  & SS (on-axis) 
  & B2 ($1.5^\circ$) 
  & GROUND ($6.0^\circ$) \\
\multicolumn{2}{l}{} 
  & $\Delta m^2_{41} = 5\,\text{eV}^2$ 
  & $\Delta m^2_{41} = 2\,\text{eV}^2$ 
  & $\Delta m^2_{41} = 1\,\text{eV}^2$ \\
\hline
\multirow{2}{*}{Intrinsic} 
  & $\nu_e$ beam         & 29738.3  & 15372.2  & 6089.58 \\
  & $\bar{\nu}_e$ beam   & 1371.01  & 1171.46  & 508.08  \\
\hline
\multirow{2}{*}{mis-ID}    
  & $\nu_\mu$ CC         & 37369.6\,(18684.8) & 9342.77\,(4471.39) & 2190.43\,(1095.2) \\
  & $\nu_\mu$ NC         & 6253.51\,(12507.0) & 1641.8\,(3283.6)   & 445.1\,(890.2)    \\
\hline
\textbf{Signal}            
  & $\nu_\mu \to \nu_e$  & \textbf{2858.12} & \textbf{549.47} & \textbf{90.28} \\
\hline\hline
\end{tabular}
\end{table}

% TABLE 2: nu_mu disappearance channel
\begin{table}
\centering
\caption{Expected background and signal event counts for the $\nu_\mu$ \textbf{disappearance} channel, for 10 ton-year exposure at each detector location. Events are generated with $\sin^2 2\theta_{\mu e} = 0.001$.}
\label{tab:disappearance}
\begin{tabular}{llccc}
\hline\hline
\multicolumn{2}{l}{}
  & SS (on-axis)
  & B2 ($1.5^\circ$)
  & GROUND ($6.0^\circ$) \\
\multicolumn{2}{l}{}
  & $\Delta m^2_{41} = 5\,\text{eV}^2$
  & $\Delta m^2_{41} = 2\,\text{eV}^2$
  & $\Delta m^2_{41} = 1\,\text{eV}^2$ \\
\hline
Intrinsic & $\bar{\nu}_\mu \to \bar{\nu}_\mu$ & 50151.9 & 21877.8 & 5610.75 \\
\hline
\textbf{Signal} & $\nu_\mu \to \nu_\mu$ & $\mathbf{3.55 \times 10^6}$ & \textbf{887563} & \textbf{208091} \\
\hline\hline
\end{tabular}
\end{table}

\newpage
\noindent
{\it Role of the $\nu_\mu$ disappearance channel} \\

\begin{figure}[H]
\begin{center}  
\includegraphics[width=82mm, height=65mm]{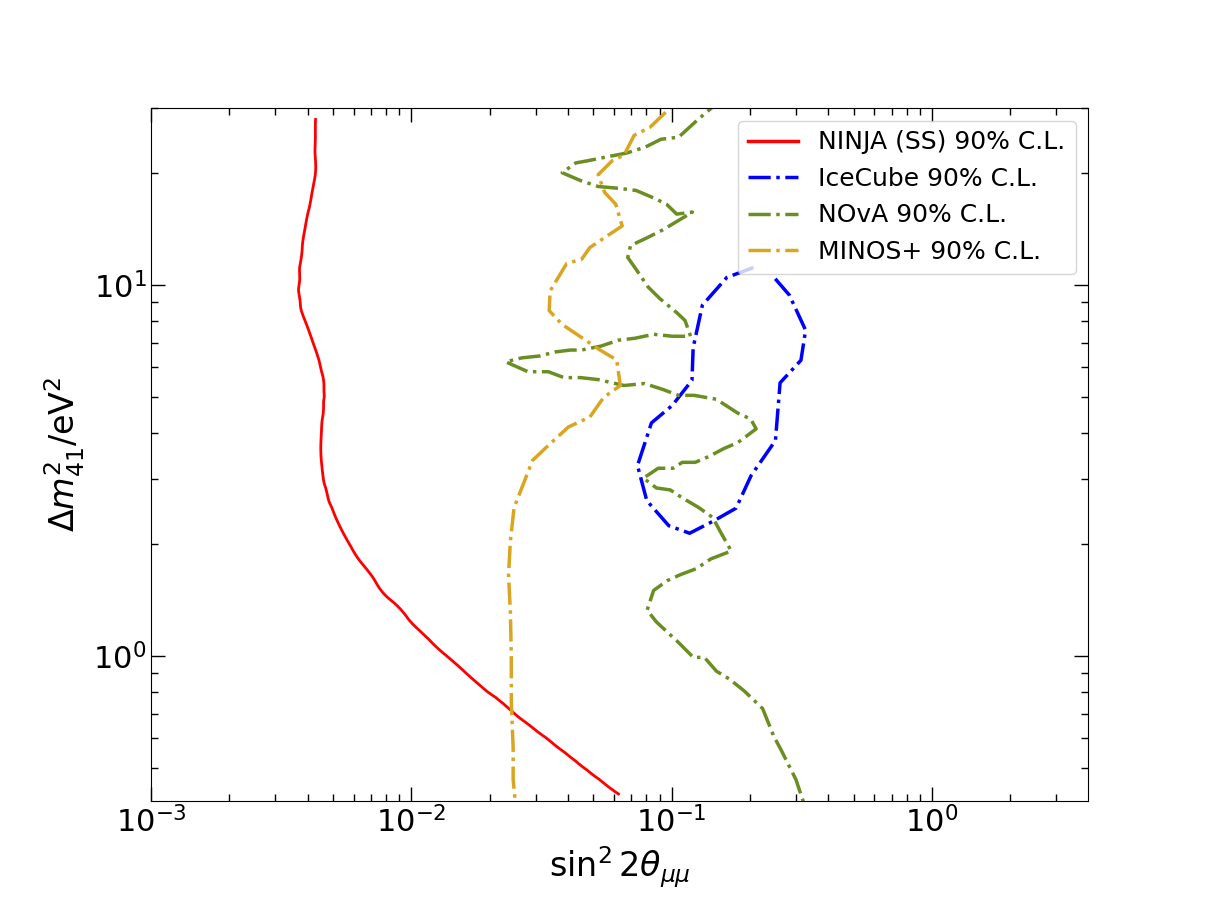}
\caption{Comparison of exclusion limits with at the SS floor for 10 ton-year exposure using only $\nu_\mu\to\nu_\mu$ disappearance channel. }
\label{fig_numu}
\end{center}
\end{figure}

The $\nu_\mu$ disappearance channel plays a major role in constraining the sterile neutrino parameter space in the present analysis. Owing to the large $\nu_\mu$ flux at J-PARC and high statistics achievable at short baselines, this channel provides strong sensitivity to the eV-scale sterile neutrinos. In Fig. \ref{fig_numu}, we show the exclusion limit in the $\sin^22\theta_{\mu\mu} - \Delta m^2_{41}$ plane obtained using only the $\nu_\mu \rightarrow \nu_\mu$ disappearance channel at the SS floor, and compare with existing constraints from IceCube~\cite{IceCube:2024nle}, NO$\nu$A~\cite{NOvA:2024imi}, and
MINOS+~\cite{MINOS:2020iqj}. We show that the $\nu_\mu$ disappearance channel is the dominant source of sensitivity in our analysis and largely determines the overall exclusion reach in the $\sin^2 2\theta_{\mu\mu}$--$\Delta m^2_{41}$ plane.

\noindent
{\it Role of the $\nu_e$ disappearance channel} \\

\begin{figure}[H]
\begin{center}  
\includegraphics[width=80mm, height=65mm]{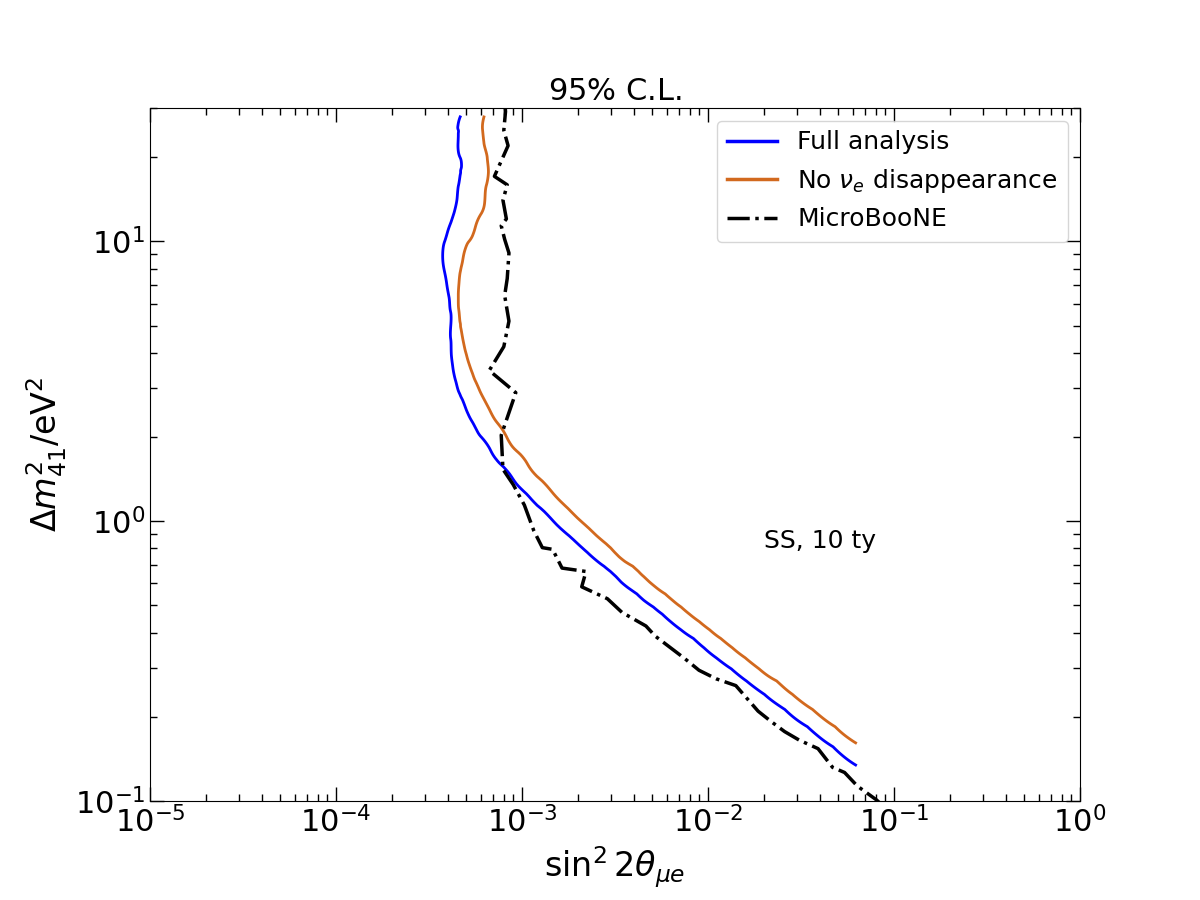}
\caption{Comparison of exclusion limits at the SS floor for 10 ton-year exposure with (solid blue) and without (solid orange) the $\nu_e\to\nu_e$ disappearance channel. }
\label{fig_nue}
\end{center}
\end{figure}

Although the main sensitivity of an accelerator based neutrino experiment comes from the $\nu_e$ appearance channel and $\nu_\mu$ disappearance channel, in principle one can also include the $\nu_e$ disappearance channel in analyzing experiment's sensitivity to the sterile neutrinos. Generally, this channel is used as an intrinsic beam background for the $\nu_e$ appearance channel. Interestingly, it has been shown in Ref.~\cite{Denton:2021czb} that though MicroBooNE does not observe any electron excess in the $\nu_e$ appearance channel, it might have an excess in the $\nu_e$ disappearance channel. However, in their subsequent analysis MicroBooNE performed a joint fit taking the $\nu_e$ appearance and $\nu_e$ disappearance channel and show that three-neutrino fit still gives a better fit than four-neutrino fit at $1\sigma$~\cite{MicroBooNE:2022sdp}. In order to show the additional constraining  of the $\nu_e$ disappearance channel, in Fig.~\ref{fig_nue} we show a comparison between exclusion curves obtained with (solid blue) and without (solid orange) this channel, for a 10 ton-year exposure at the SS floor. Including the $\nu_e$ disappearance channel yields a modest strengthening of the exclusion, demonstrating the value of combining both appearance and disappearance information in the 3+1 analysis. 

\noindent
{\it Role of backgrounds} \\

\begin{figure}[H]
\centering
\begin{subfigure}[t]{0.49\textwidth}
    \centering
    \includegraphics[width=1.1\textwidth,height=64mm]{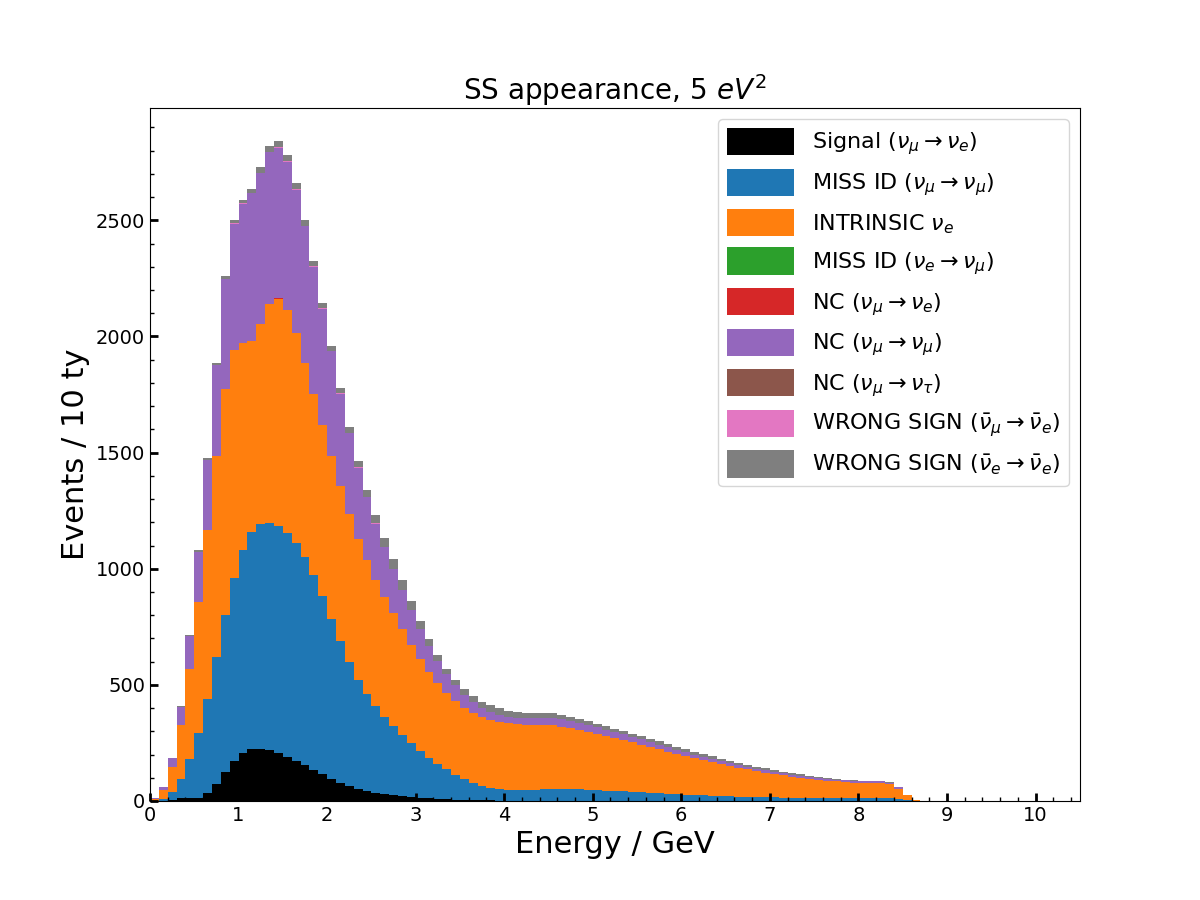}
    \caption{Event spectra for signal and background for SS floor.}
    \label{fig:sens_events}
\end{subfigure}%
\hfill
\begin{subfigure}[t]{0.49\textwidth}
    \centering
    \includegraphics[width=\textwidth, height=64mm]{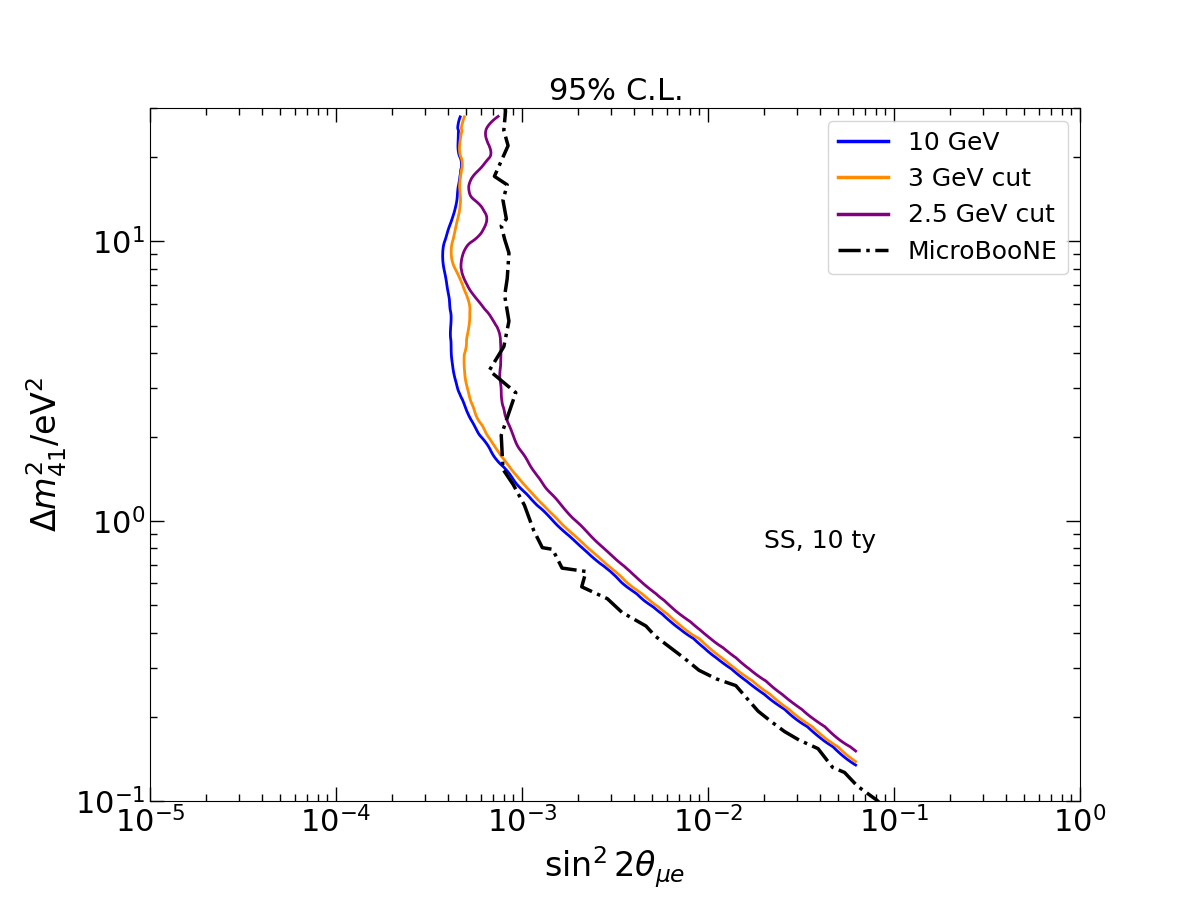}
    \caption{Sensitivity without (10 GeV) and with energy cuts.}
    \label{fig:sens_cuts}
\end{subfigure}
\caption{Impact of energy selection on sterile neutrino sensitivity at the SS floor.}
\label{fig_event}
\end{figure}

To understand which backgrounds limit the sensitivity the most, we examine the signal and background spectra at the SS floor. The left panel of Fig.~\ref{fig_event} presents the energy spectra for both signal and background events, assuming $\Delta m^2_{41} = 5~eV^2$ and $\sin^2 2\theta_{\mu e} = 0.001$ for the $\nu_e$ appearance channel. This panel illustrates that above 2~GeV, the frequency of signal events declines sharply, whereas the number of background events remains quite significant. Consequently, implementing an upper energy cut around 2~GeV could enhance sensitivity by improving the signal-to-background ratio. 
To investigate this, the right panel of Fig.~\ref{fig_event} displays the sensitivity for various energy cutoffs: the blue curve represents the scenario without an energy cut (up to 10~GeV), whereas the purple and orange curves correspond to upper energy cuts at 2~GeV and 3.5~GeV, respectively. Although there is not a significant number of \(\nu_e\) events above 3.5~GeV, results indicate that applying energy cuts leads to a deterioration in sensitivity compared to the no-cut case. The reason for this is the fact that our sensitivity is primarily influenced by the combination of the \(\nu_e\) appearance and \(\nu_\mu\) disappearance channels. Therefore, although the number of \(\nu_e\) events from the appearance channel is not large, a substantial number of \(\nu_\mu\) events from the disappearance channel, as well as some \(\nu_e\) appearance events occur above 3.5~GeV which contributes significantly to the overall sensitivity. As a result, applying an upper cut at 3.5~GeV reduces the total event statistics and leads to a deterioration in sensitivity. \\

\noindent
{\it Role of detector efficiencies} \\

\begin{figure}[H]
\begin{center}  
\includegraphics[width=80mm, height=65mm]{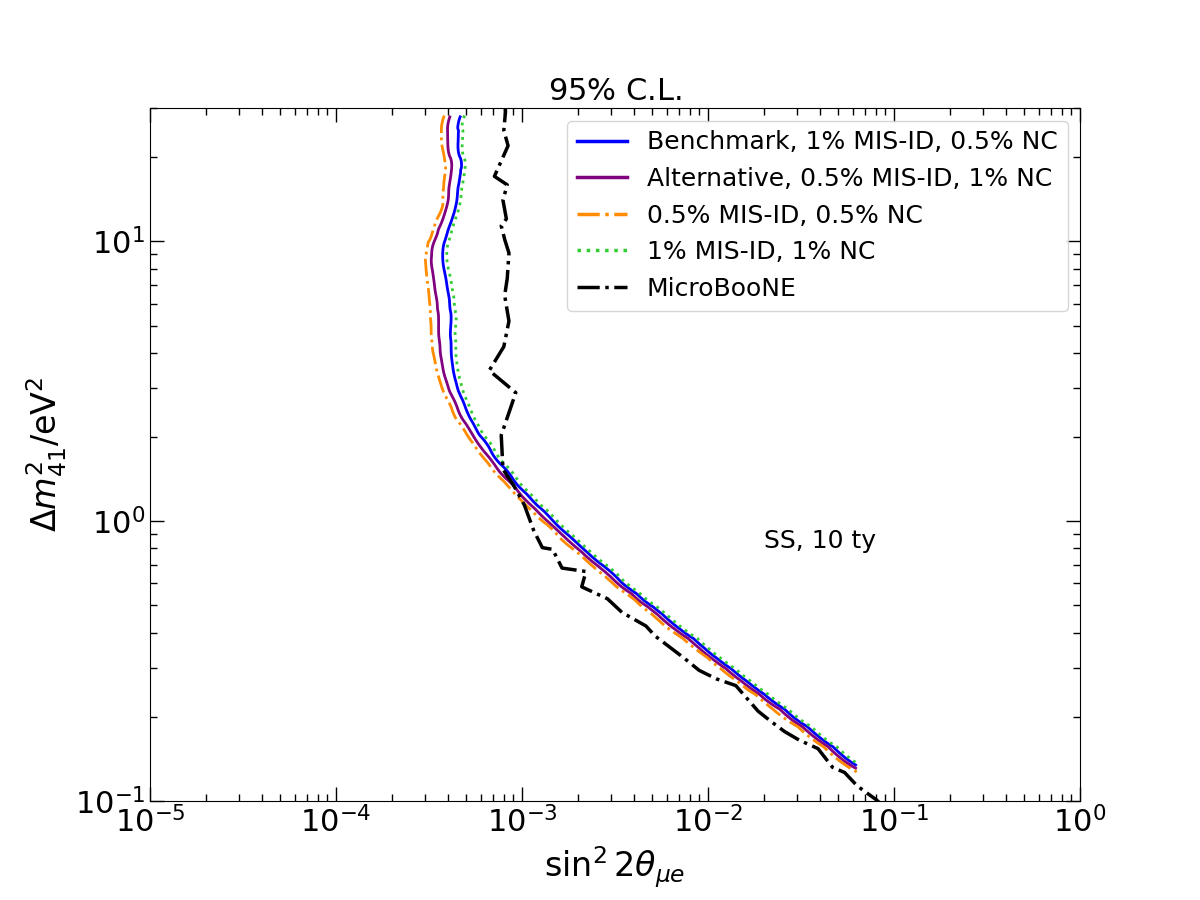}
\caption{Comparison of the impact of varying $\nu_\mu$ CC and NC misidentification backgrounds.}
\label{fig:fig_eff}
\end{center}
\end{figure}

Next, we investigate the impact of detector efficiencies on the sensitivity. To assess this effect, we compare the benchmark and alternative sets of efficiencies for $\nu_\mu$ CC mis-ID and $\nu_\mu$ NC mis-ID backgrounds. In future runs of the NINJA-like detector, we anticipate improved muon identification at the energies at which the SS floor operates, leading to a reduction in the $\nu_\mu$ CC mis-ID background. In contrast, distinguishing electron signal events from the decay products of $\pi^0$ mesons becomes more challenging in this energy regime, which may lead to a reduced efficiency for NC background rejection. Fig.~\ref{fig:fig_eff} illustrates how the sensitivity of NINJA changes with respect to the change in efficiency of the mentioned backgrounds. To better understand the relative impact of each background, we present the sensitivity obtained by modifying only one efficiency at a time. From this comparison, we find that improving the muon CC misidentification efficiency from 1\% to 0.5\% has a larger effect on sensitivity than worsening the NC background efficiency from 0.5\% to 1\%. This is primarily because muon events are abundant in CC interactions, so even a small misidentification rate leads to a significant background in the electron appearance channel. In contrast, NC events occur less frequently. Consequently, a modest change in NC rejection has a smaller impact on the overall sensitivity. Finally, we note that more realistic efficiency estimates are expected to become available from the upcoming RUN~6 iron ECC data, which will be incorporated in future analyses. \\

\noindent
{\it Role of systematic uncertainties} \\

\begin{figure}[H]
\centering
\begin{subfigure}[t]{0.49\textwidth}
    \centering
    \includegraphics[width=1\textwidth,height=64mm]{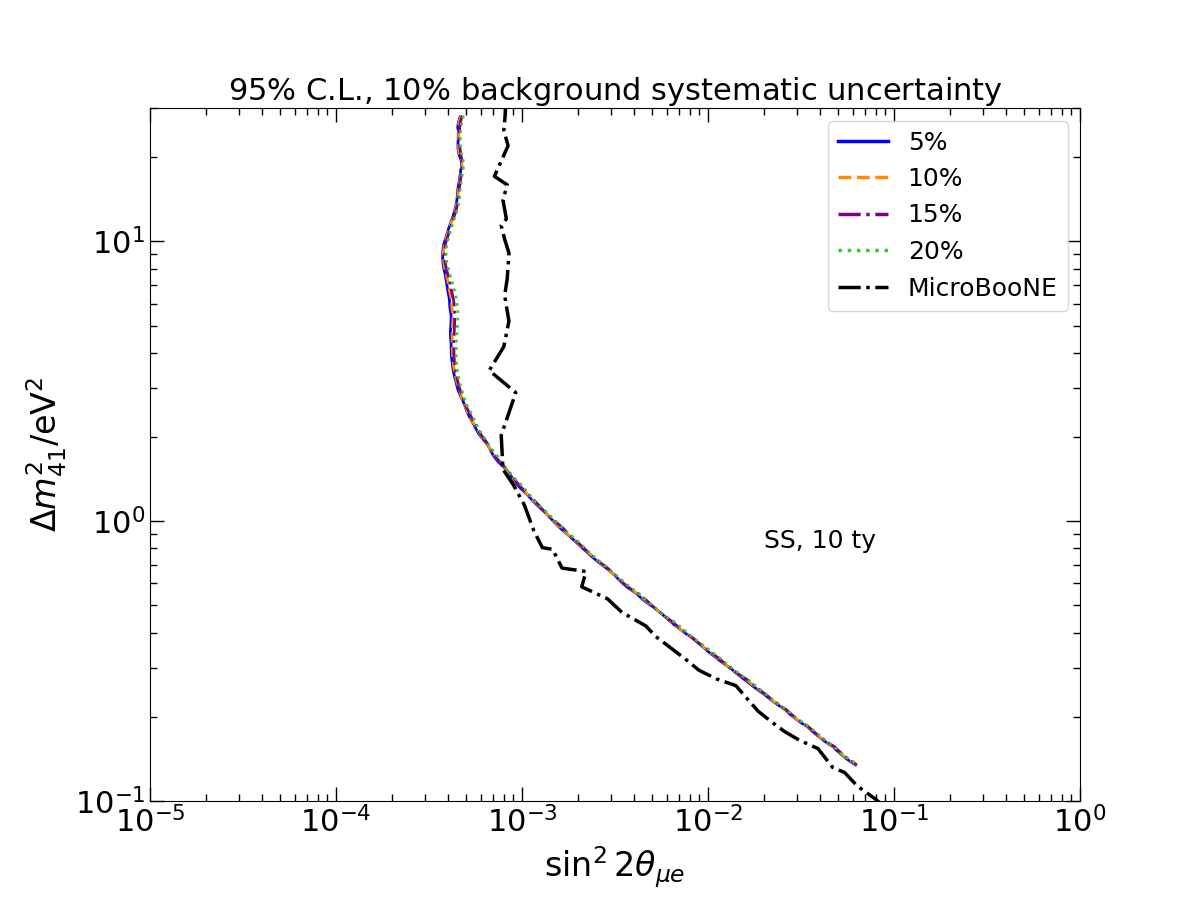}
    \caption{Sensitivity for different levels of systematic uncertainty on the signal events, with the systematics errors for the background fixed at 10\%. Only normalization error is considered.}
    \label{fig:fig_sys}
\end{subfigure}%
\hfill
\begin{subfigure}[t]{0.49\textwidth}
    \centering
    \includegraphics[width=\textwidth, height=64mm]{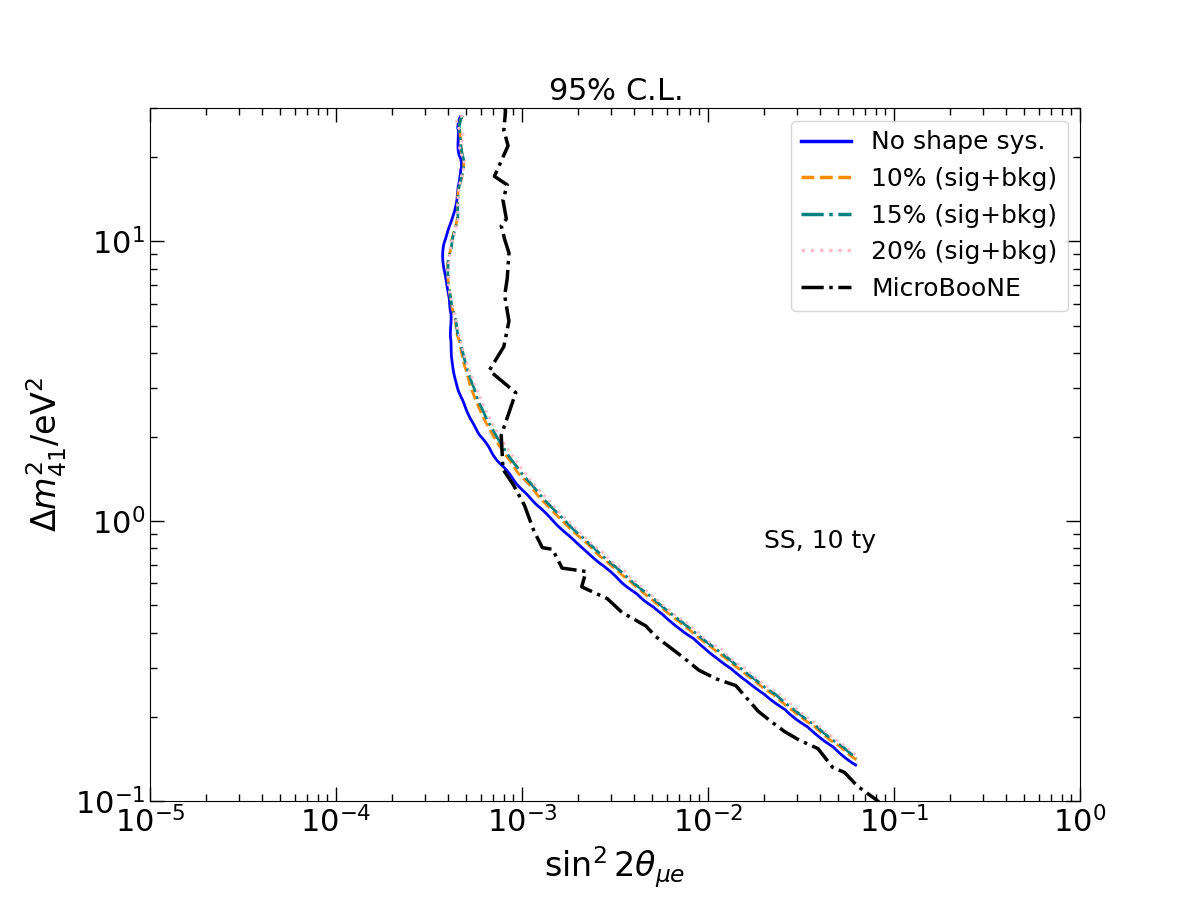}
    \caption{Sensitivity for different levels of systematic uncertainty for both signal and background events. Both normalization and shape error is considered. }
    \label{fig:fig_sys_shape}
\end{subfigure}
\caption{Impact of systematic uncertainty on sterile neutrino sensitivity at the SS floor.}
\label{fig_sys}
\end{figure}

Since NINJA does not have a near detector, the neutrino flux from the J-PARC source cannot be measured very accurately. As a result, the experiment may be subject to sizable normalization uncertainties. To quantify this effect, Fig.~\ref{fig:fig_sys} shows the sensitivity curves for four different assumptions of the overall normalization error for the SS floor. Interestingly, we observe only a mild deterioration in sensitivity when increasing the systematic uncertainty on the signal events from 5\% to 20\%. This robustness with respect to normalization uncertainties reflects the short-baseline nature of the measurement, where high event rates are achievable even with modest exposures. Next, to illustrate the effect of systematics in the spectral shape, in Fig.~\ref{fig:fig_sys_shape}, Fig.~\ref{fig:fig_sys_shape} shows how a shape error changes the sensitivity. To introduce the shape error, we varied the events $N$ in each energy bin $i$ in the following way:
\begin{equation}
    N_i \rightarrow N_i [( \xi \times \zeta(E_i^{\prime} - \bar{E^{\prime}})/(E^{\prime}_{\rm max}-E^{\prime}_{\rm min})],
    \label{error}
\end{equation}
where $\zeta$'s are pull variables associated with the shape uncertainty, $\xi$. Here, $E^{\prime}_{\rm max}$ and $E^{\prime}_{\rm min}$ denote the maximum and minimum energies of the event spectrum, and $\bar{E}^{\prime} = \frac{1}{2}(E^{\prime}_{\rm max} + E^{\prime}_{\rm min})$ represents the midpoint of the energy interval, and $E_i^{\prime}$ is the reconstructed energy of the $i^{\rm th}$ bin. This parametrization of the shape error is standard in GLoBES and is therefore used in studies based on this software. For example, the results shown in the conceptual design report of ESSnuSB \cite{Alekou:2022emd} also use this parametrization for shape error. Our results show that introducing a shape error slightly shifts the sensitivity in comparison to the case without it. However, when the shape error is varied from 10\% to 20\%, the effect is negligible. \\

\noindent
{\it Role of energy resolution} \\

\begin{figure}[h]
\begin{center}  
\includegraphics[width=80mm, height=65mm]{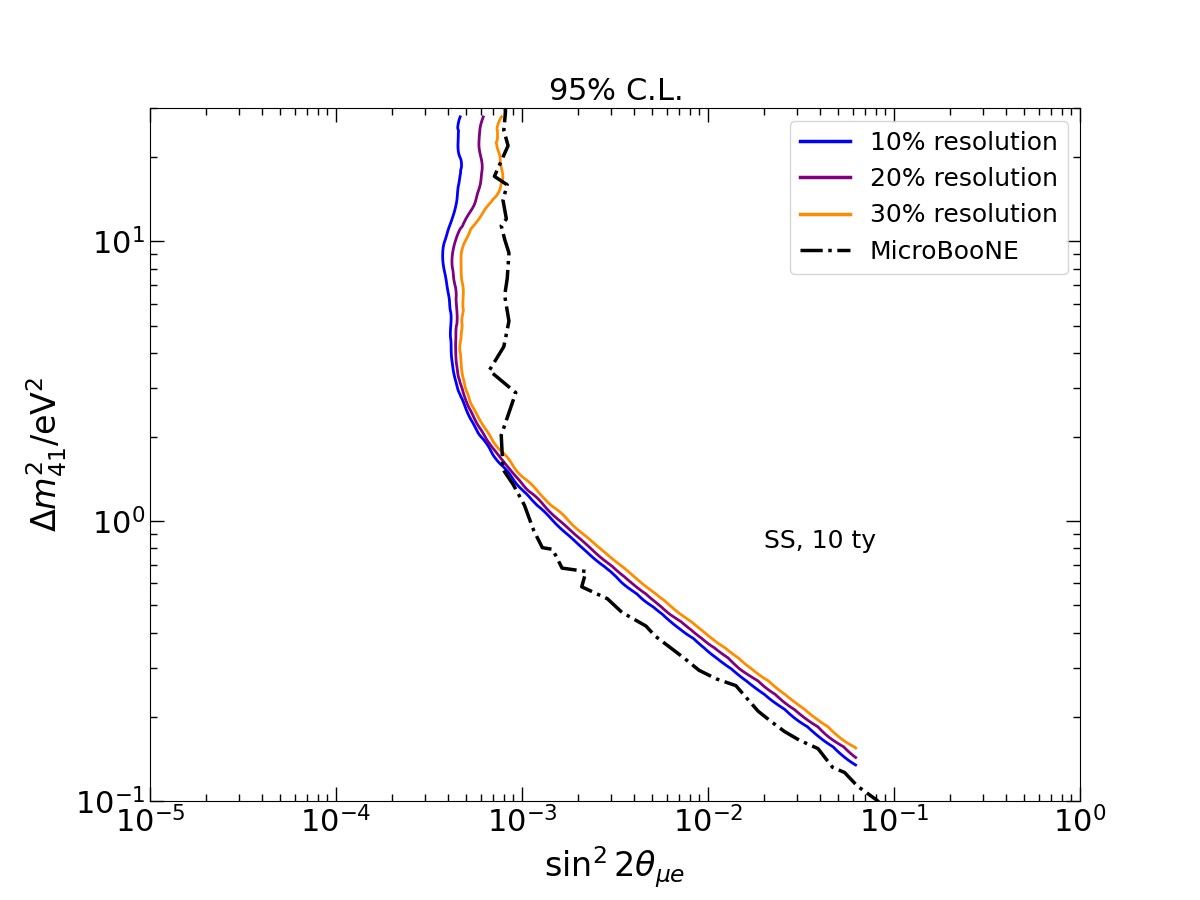}
 \caption{Impact of the assumed neutrino energy resolution on the sensitivity.}
\label{fig:resolution}
\end{center}
\end{figure}

In the absence of a dedicated detector simulation for a possible future NINJA configuration, the neutrino energy reconstruction is modeled using a Gaussian smearing with a fixed fractional energy resolution. In the baseline analysis, we adopt a 10\% energy resolution, which should be regarded as an illustrative benchmark rather than a realistic expectation for an ECC-only detector. In practice, an ECC based detector is expected to achieve an energy resolution of about 20-30\%, while improved performance is possible when complemented by an external muon spectrometer. To assess the impact of a more realistic energy resolution, we examine how the projected sensitivity would change if the energy resolution were degraded to 20\% or 30\%. As shown in Fig.~\ref{fig:resolution}, the effect on the sensitivity is minimal, particularly in the region of interest around $\Delta m^2_{41} \sim 5~\mathrm{eV}^2$. This indicates that the sterile neutrino sensitivity of a NINJA-like detector is insensitive to the assumed energy resolution within the range considered. \\

\noindent
{\it Sensitivity at SS+B2 floor} \\

\begin{figure}[h!]
\begin{center}  
\includegraphics[width=80mm, height=65mm]{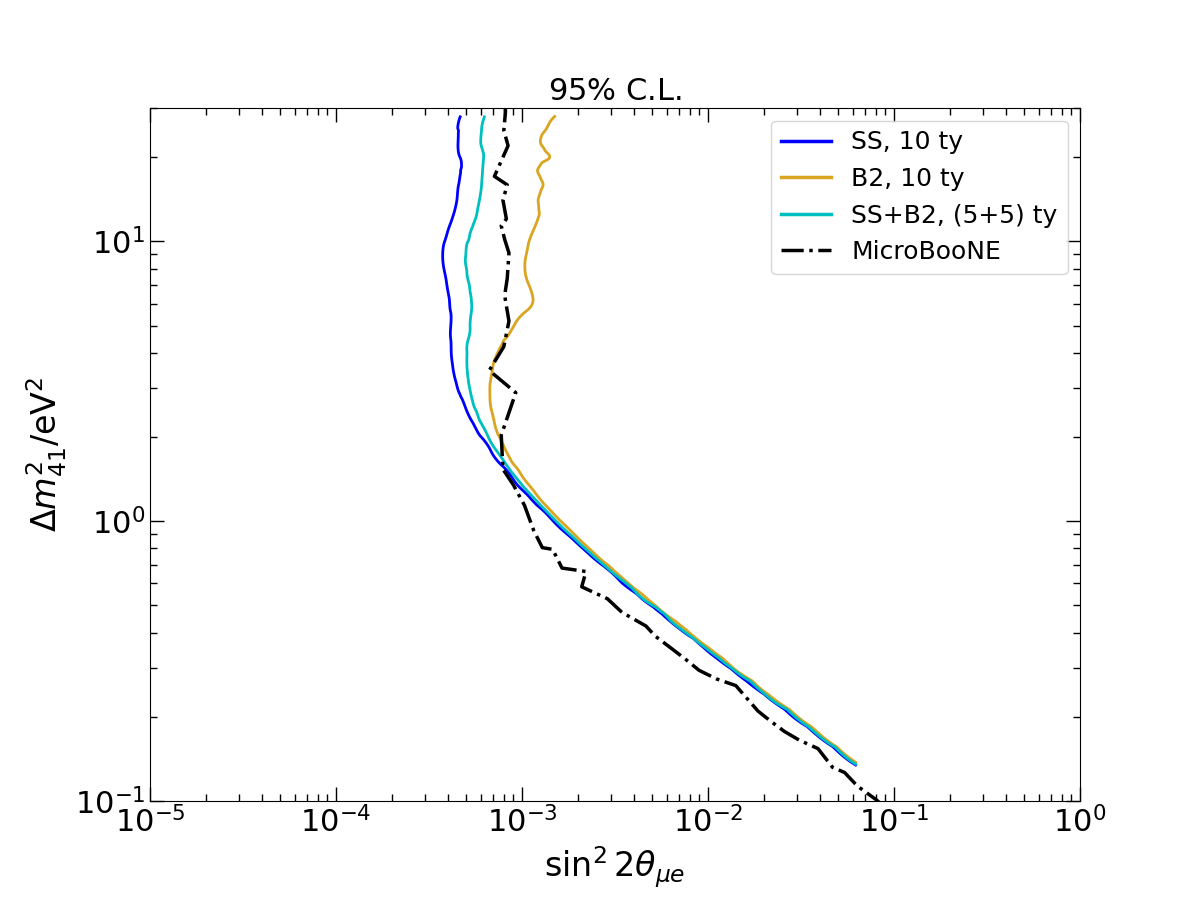}
\caption{Bounds on the sterile neutrino mixing parameters for three configurations: 10 ty exposure at the SS floor, 10 ty at the B2 floor, and a combined analysis with 5 ty exposure at each location (SS+B2). }
\label{fig_combined}
\end{center}
\end{figure}

In the Fig.~\ref{fig_combined}, we examine the combined sensitivity of the possible future NINJA detector, assuming a split exposure: 5~ton-years at the SS floor and 5~ton-years at the B2 floor. This configuration serves two key purposes. First, utilizing both off-axis (B2) and on-axis (SS) fluxes, NINJA gains sensitivity across a broader range of mass-squared differences, $\Delta m^2_{41}$. Second, if the SS floor does not provide sufficient space for the full detector volume, the detector can physically be distributed between both locations. From the figure, we observe that the combined B2+SS setup indeed achieves sensitivity over a wide range of $\Delta m^2_{41}$. However, the overall sensitivity lies between that of the individual SS-only and B2-only configurations. For this combined setup, we explored what happens if one assumes correlated systematics between the SS and the B2 floor, as compared to the case with uncorrelated systematics. However, we found that both scenarios yield almost similar results, implying the statistics-dominant nature of the sensitivity.

\section{Conclusion}
\label{sec:conclusion}
In this paper, we have demonstrated the potential of the NINJA experiment to probe eV-scale sterile neutrinos in its future runs. The distance between the NINJA detector and the neutrino source, together with the energy of the neutrinos from J-PARC, provides an excellent opportunity to search for light sterile neutrinos at short baselines. In our analysis, we consider a lead-based target and adopt a simplified detector response in which constant benchmark selection efficiencies are applied. These efficiencies are interpreted as performance requirements motivated by the demonstrated capabilities of emulsion cloud chamber technology and other accelerator-based experiments, rather than as predictions based on a detailed detector simulation. 

Our results show that with a 10~ton-year exposure, the SS-floor flux provides strong constraints on the 3+1 sterile-neutrino parameter space, reaching sensitivity comparable to existing accelerator-based limits. Even a reduced exposure of 4~ton-years on the SS floor yields competitive sensitivity. In contrast, bounds obtained from the B2 (GROUND) floor flux are comparable (weaker) or, in some regions, complementary to existing constraints. Imposing an upper energy cut does not improve the SS-floor sensitivity, as the signal-to-background ratio remains higher at higher energies.
We have demonstrated the importance of combining appearance and disappearance channels in the sensitivity analysis, resulting in strengthened bounds on sterile-neutrino parameters. However, the overall sensitivity is driven primarily by the $\nu_\mu$ disappearance channel. Variations of the efficiencies for the dominant backgrounds show that muon charge current misidentification has a noticeable impact on the sensitivity, while misidentification due to muon neutral current backgrounds lead to only modest changes. The exclusion reach is dominated by statistical uncertainties, whereas the effect of overall systematics is found to be negligible. This holds both for the normalization uncertainty and for uncertainties in the spectral shape: introducing a shape error produces only a slight shift in sensitivity relative to the case with no shape error, and varying its size between 10\% and 20\% has a negligible further effect.  Placing a near detector close to the source may help to have a better knowledge on the flux uncertainty and therefore on systematics.  However, installing an additional detector at a shorter baseline is not considered feasible within the existing J-PARC neutrino beamline configuration, because of the  constraints coming from the beamline infrastructure, decay volume, and radiation shielding around the beam dump. 
We have also studied the robustness of our results with respect to the assumed neutrino energy resolution. Varying the resolution between 10\%, 20\% and 30\% leads to only modest changes in the exclusion sensitivity in the region of interest, indicating that our conclusions are not strongly dependent on the precise choice of energy resolution within a reasonable range. Our results also show that a split exposure of 5~ton-years at the SS floor and 5~ton-years at the B2 floor can provide sensitivity for a wide range of $\Delta m^2_{41}$. However, in this case sensitivity is weaker than the sensitivity of the SS floor only, with the exposure of 10~ton-years.  Note that though in this present work we have considered only neutrino mode, adding antineutrino is not expected to effect the sensitivity. As this is mainly two flavour case, there exist no parameter degeneracy which might benefit from adding antineutrinos. On the other hand, replacing neutrino runtime by antineutrino might cause a reduction in the sensitivity due to its lower flux and smaller cross-section.

Finally, we point out that if sterile neutrinos exist in nature, any measurement of the intrinsic $\nu_e$ cross section in NINJA would be contaminated by $\nu_e$ appearance from $\nu_\mu \rightarrow \nu_e$ oscillations. One possible way to address this issue is to extract cross sections in energy regions where the sterile-oscillation probability is negligible for a given flux configuration. For example, for the SS-floor flux, this corresponds to energies above approximately 3.5~GeV, where the expected number of $\nu_e$ appearance events is very small. However, this region is also characterized by a large contribution from the $\nu_e$ disappearance channel, making an accurate flux
determination essential to disentangle these effects.

\section*{Appendix A: Effect of baseline smearing, misidentification background and energy resolution}

In the nominal analysis, the neutrino oscillation probabilities are evaluated at a fixed baseline of 280\,m, corresponding to the distance between the production target and the detector. In reality, the parent mesons propagate along the decay pipe before decaying, so neutrinos are produced at different positions and consequently travel different distances to the detector. To investigate the effect of this baseline smearing, we consider the 96\,m decay pipe and approximate the distribution of neutrino production positions using the decay probability of the parent pions.

For a neutrino produced at a distance $x$ downstream of the target, the baseline is $L(x)=L_0-x$, where $L_0=280\,\mathrm{m}$. We assume an exponential pion-decay distribution, proportional to $\exp[-x/\lambda_\pi(E_\nu)]$, with the pion decay length given by $\lambda_\pi(E_\nu)=c\tau_\pi p_\pi(E_\nu)/m_\pi$. Here, $c$ is the speed of light, $\tau_\pi$ is the proper lifetime of the pion, $m_\pi$ is the mass of the pion and $p_\pi$ is the parent-pion momentum, estimated from the neutrino energy $E_\nu$ using forward decay kinematics $\pi\rightarrow\mu\nu$. The oscillation probability is then averaged over the possible production positions according to

\begin{equation}
\overline{P}_{\alpha\beta}(E_\nu)=
\frac{
\displaystyle\int_0^{D} dx\,
e^{-x/\lambda_\pi(E_\nu)}
P_{\alpha\beta}(E_\nu,L_0-x)
}{
\displaystyle\int_0^{D} dx\,
e^{-x/\lambda_\pi(E_\nu)}
},
\label{eq:baseline_avg}
\end{equation}

where $D=96\,\mathrm{m}$ is the decay-pipe length.. The denominator ensures that the decay-position distribution is normalized over the finite decay volume.

We implement this averaging directly in the GLoBES probability engine by evaluating the full oscillation probability matrix at discrete positions along the decay pipe and combining the results with the corresponding exponential weights. The averaged probabilities are subsequently used in the event-rate calculation and the $\chi^2$ sensitivity analysis, while all detector assumptions are kept unchanged.

\begin{figure}[H]
\begin{center}
\includegraphics[width=80mm,height=65mm]{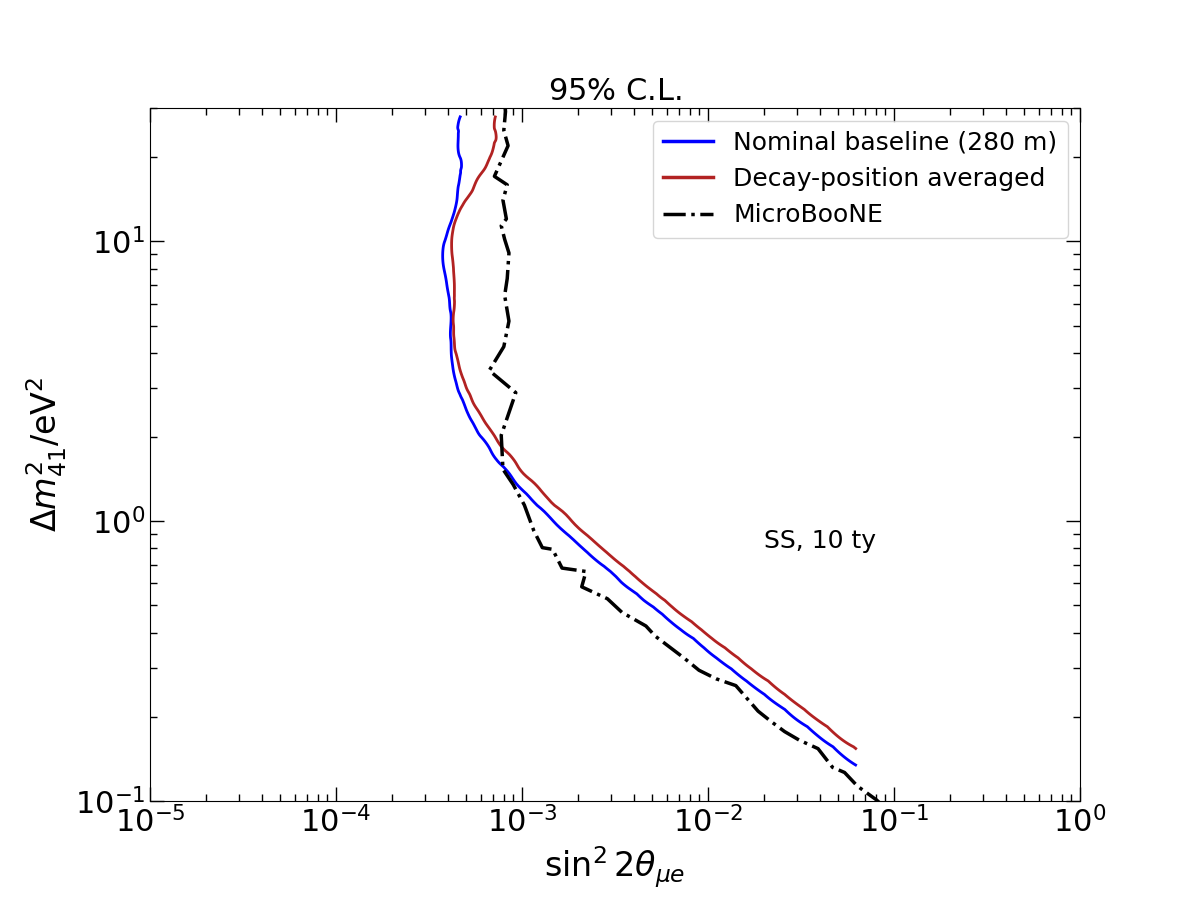}
\caption{Comparison of the 95\% C.L. exclusion limits at the SS location for a 10 ton-year exposure, obtained using the nominal fixed baseline of 280\,m and the decay-position-averaged oscillation probabilities. The MicroBooNE exclusion limit is shown for comparison. The benchmark mis-ID probabilities are used in both calculations.}
\label{fig_avg}
\end{center}
\end{figure}

Figure~\ref{fig_avg} compares the resulting 95\% C.L. exclusion contour with the original fixed-baseline calculation for the SS location and an exposure of 10 ton-years. The inclusion of baseline smearing mainly affects the region with larger $\Delta m^2_{41}$ values. For example, above $\Delta m_{41}^{2}\sim 10\,\mathrm{eV}^{2}$, the difference becomes larger, with the averaged calculation yielding weaker exclusion limits. This is expected because the oscillation phase varies more rapidly with the baseline at larger $\Delta m_{41}^{2}$, leading to stronger averaging of the oscillatory features across the decay volume. However, around $\Delta m_{41}^{2}\sim 5\,\mathrm{eV}^{2}$, where the experiment can achieve its strongest sensitivity, the two contours nearly coincide, making the impact of baseline smearing negligible in this region.

The decay-position averaging uses an approximate pion-decay distribution, with the parent-pion momentum estimated from two-body decay kinematics. This approximation is most appropriate for the pion-originated $\nu_\mu$ component, which dominates the projected sensitivity. Other contributions, including intrinsic $\nu_e$ from muon and kaon decays, may have different production-position distributions that are not separately modeled in this calculation.

\begin{figure}[H]
\begin{center}  
\includegraphics[width=80mm, height=65mm]{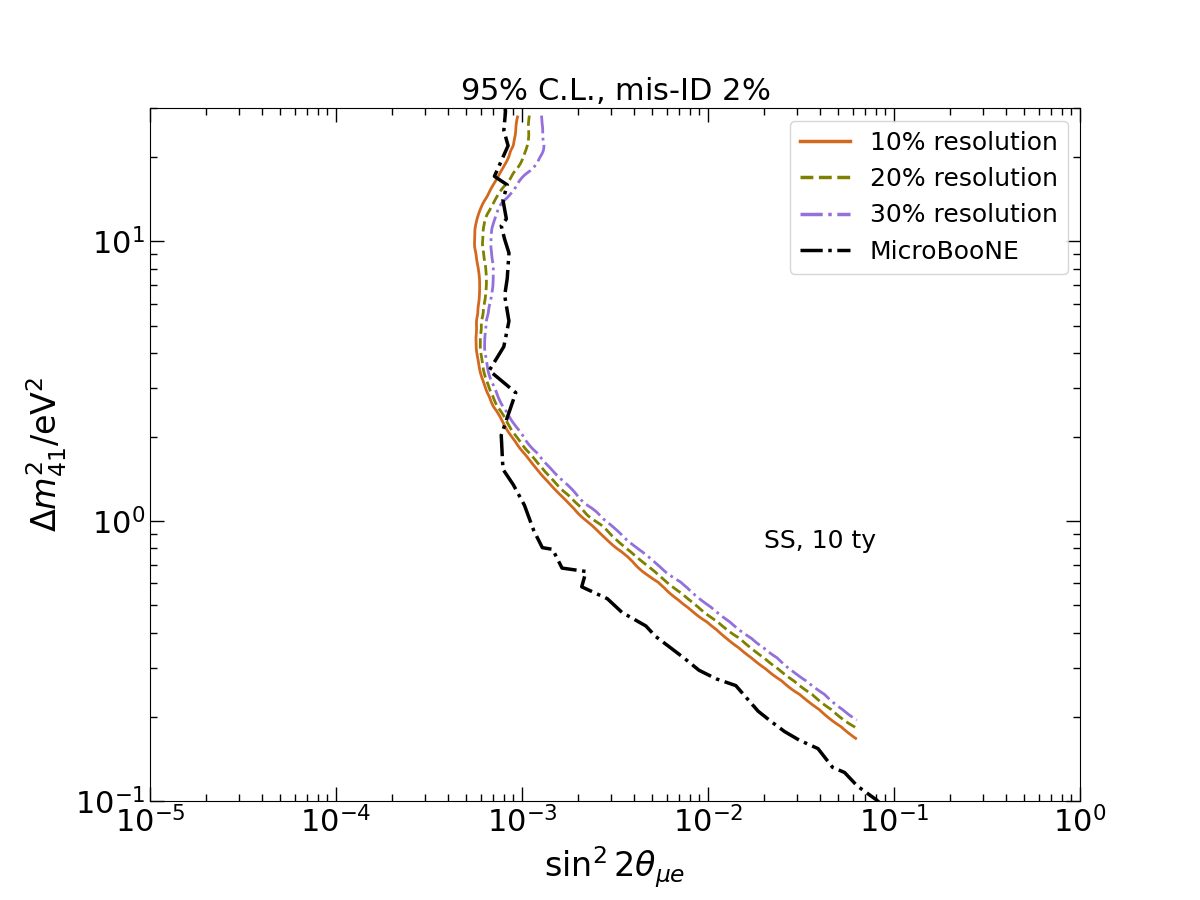}
\caption{Comparison of exclusion limits at the SS floor with different energy resolutions, for 10 ton-year exposure, obtained using the decay-position-averaged oscillation probabilities and an alternative 2\% mis-ID efficiency. }
\label{fig_avg_resolution}
\end{center}
\end{figure}

To further investigate the impact of baseline smearing under more conservative detector assumptions, we repeated the energy-resolution study using a 2\% $\nu_\mu$ CC misidentification probability and the decay-position-averaged oscillation probabilities. The resulting exclusion contours for different energy resolutions are shown in Fig.~\ref{fig_avg_resolution}. The sensitivity remains only weakly dependent on the assumed energy resolution, particularly around $\Delta m_{41}^{2}\sim 5\,\mathrm{eV}^{2}$. This confirms that the limited impact of energy-resolution variations observed in the original analysis persists even when baseline smearing is included. Note that with worsening of mis-ID background, the sensitivity reduces, as already shown in Fig.~\ref{fig:fig_eff}. 

\section*{Appendix B: Comparison with SBN bound}

\begin{figure}[H]
\begin{center}  
\includegraphics[width=76mm, height=67mm]{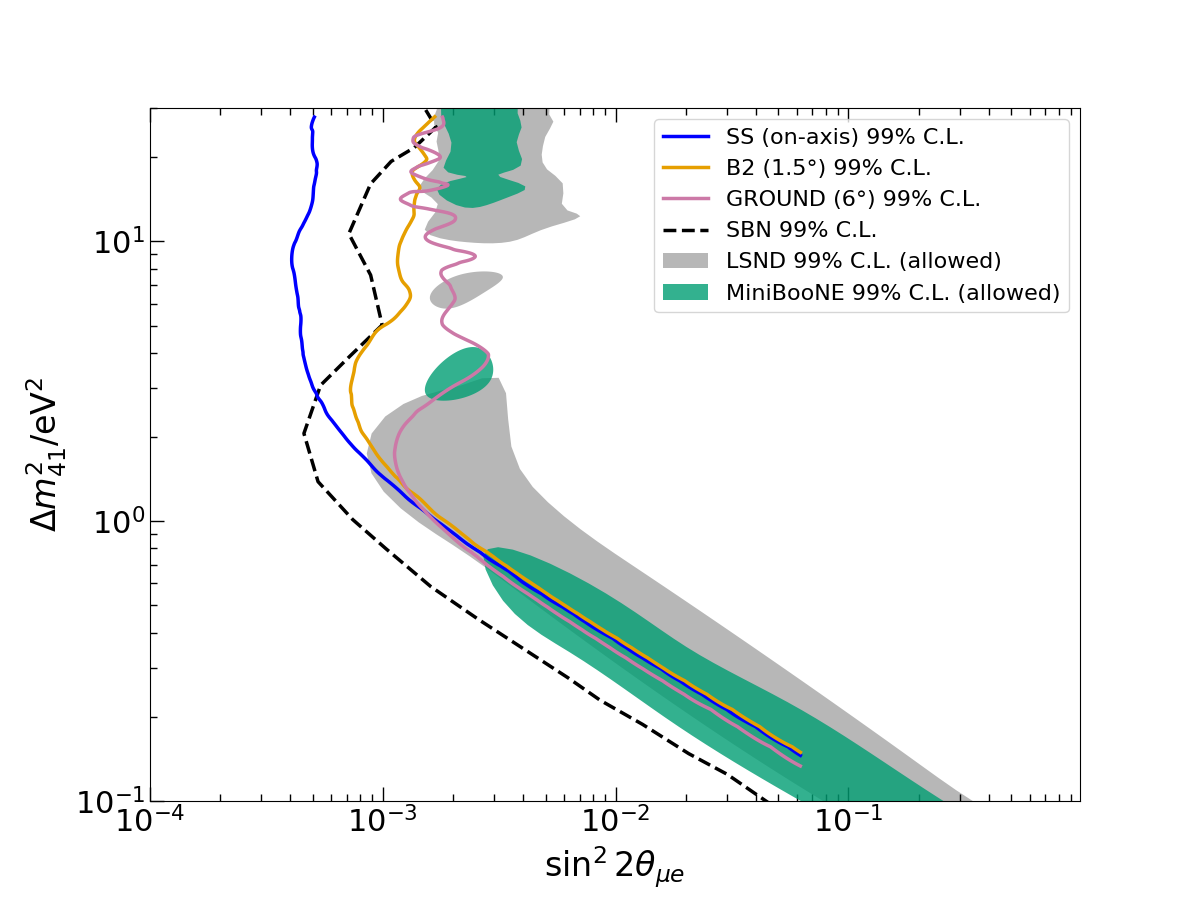}
 \caption{Comparison of the sensitivity for three detector locations with 10 ton-year
exposure with allowed region on $99\%$ C.L. from MiniBooNE \cite{MiniBooNE:2020pnu}, LSND \cite{LSND:2001aii} and estimated SBN bound \cite{SBND:2025lha}.}
\label{fig:fig_sbn}
\end{center}
\end{figure}

Fig.~\ref{fig:fig_sbn} shows the comparison of the sensitivity at 99\% C.L. between NINJA and SBN facility \cite{SBND:2025lha} which includes the combined  estimated sensitivity of MicroBooNE, ICARUS and SBND. The comparison indicates that NINJA could achieve better sensitivity in certain ranges of $\Delta m^2_{41}$.

\section*{Acknowledgements}

We thank the NINJA collaboration for support, discussion and encouragement throughout this work. DB and MG thank Naoki Otani for useful comments. This work has been funded in part by Ministry of Science and Education of Republic of Croatia grant No. PK.1.1.10.0002, Swiss National Science Foundation (SNSF) and Croatian Science Foundation (HRZZ) under grant MAPS IZ11Z0$\_$230193 and European Union under the NextGenerationEU Programme. Views and opinions expressed are, however, those of the author(s) only and do not necessarily reflect those of the European Union. Neither the European Union nor the granting authority can be held responsible for them.

%%%%%%%%%%%%%%%%%%%%%%%%%%%%%%%%%%%%%%%%%%%%%%%%%%%%%%%%%%%%%%%%%%%%%
\bibliographystyle{JHEP}
\bibliography{reference}

\end{document}

%% file: authors.tex
\newcommand{\authorlist}{

\author[1,*]{Doris Barčot,\note[*]{Corresponding author}}
\author[2]{Tsutomu Fukuda,}
\author[1,*]{Monojit Ghosh,}
\author[1]{Leon Halić,}
\author[1]{Mahesh Jakkapu,}
\author[3]{Teppei Katori,}
\author[1]{Budimir Kliček,}
\author[2]{Masahiro Komatsu,}
\author[2]{Tomokazu Matsuo,}
\author[2]{Osamu Sato,}
\author[4]{and Atsumu Suzuki}

\affiliation[1]{Center of Excellence for Advanced Materials and Sensing Devices, Ru{\dj}er Bo\v{s}kovi\'c Institute, 10000 Zagreb, Croatia}
\affiliation[2]{Department of Physics, Nagoya University, Nagoya 464–8602, Japan}
\affiliation[3]{King’s College London, Department of Physics, Strand, London WC2R 2LS, United Kingdom}
\affiliation[4]{Kobe University, Kobe 657-8501, Japan}

\emailAdd{dbarcot@irb.hr}

}